\documentclass[a4paper,12pt]{article}

\usepackage{graphicx}
\usepackage{amsmath}
\usepackage{multirow}

\begin{document}

\begin{flushright}
preprint SHEP-06-01\\
\today
\end{flushright}
\vspace*{1.0truecm}

\begin{center}
{\large\bf Di-photon Higgs signals at the LHC in the \\[0.15cm]
Next-to-Minimal Supersymmetric Standard Model}\\
\vspace*{1.0truecm}
{\large Stefano Moretti and Shoaib Munir}\\
\vspace*{0.5truecm}
{\it High Energy Physics Group, School of Physics \& Astronomy, \\
 University of Southampton, Southampton, SO17 1BJ, UK}
\end{center}

\vspace*{2.0truecm}
\begin{center}
\begin{abstract}
\noindent
The NMSSM contains a Higgs singlet in addition to the two
Higgs doublets typical of the MSSM, thus resulting in a total of seven
physical Higgs mass states. Therefore, the phenomenology of the NMSSM Higgs sector
can vary considerably from that of the MSSM and there are good prospects
of finding in regions of the NMSSM parameter space Higgs signals 
that cannot be reproduced in the MSSM. We examined here the two-photon decay mode 
of a Higgs boson and found that up to three neutral Higgs states, heavy
and/or light, could be simultaneously observable at the LHC,
a possibility precluded to the MSSM. There are also some possibilities that
only the lightest NMSSM Higgs boson be detectable via this mode, with a mass
beyond the upper limit of the corresponding MSSM state, thus also 
allowing to distinguish between the two scenarios. However, in most of the 
NMSSM parameter space the configurations of the non-minimal model are not
very different from those arising in the minimal case.
\end{abstract}
\end{center}

\newcommand{\nn}{\nonumber}

\section{\large Introduction}

The Minimal Supersymmetric Standard Model
(MSSM) \cite{MSSMrev} is affected by the so-called `$\mu$-problem'. The MSSM Superpotential can be
written as \cite{VHiggs}
\begin{equation}
W_{\rm MSSM} = {\hat{Q}}{\hat{H}_u}{\bf{h_u}}{\hat{U}^C} 
             + {\hat{H}_d}{\hat{Q}}{\bf{h_d}}{\hat{D}^C} 
             + {\hat{H}_d}{\hat{L}}{\bf{h_e}}{\hat{E}^C} 
             + \mu{\hat{H}_u}{\hat{H}_d}.
\end{equation}
\noindent(Hereafter, hatted variables describe Superfields
while un-hatted ones stand for the corresponding scalar Superfield
components).
The last term in the above equation contains a dimensionful parameter,
$\mu$. Upon Electro-Weak Symmetry Breaking (EWSB), it 
provides a contribution to the masses of both Higgs bosons and Higgsino
fermions. Furthermore, the associated soft Supersymmetry (SUSY) breaking
term  $B\mu H_uH_d$ mixes the two Higgs doublets. Now,
the presence of $\mu$ in the Superpotential before EWSB 
indicates that its natural value would be
either 0 or the Planck mass $M_P$. On the one hand, $\mu = 0$ would mean no
mixing is actually generated between Higgs doublets at any scale and the
minimum of the Higgs potential occurs for $< H_d >
= 0$, so that one would have in turn  massless down-type fermions
and leptons after SU(2) symmetry breaking. On the other hand, $\mu \approx M_P$
would reintroduce a `fine-tuning problem' in the MSSM since the Higgs scalars
would acquire a huge contribution $\sim\mu^2$ to their squared masses (thus
spoiling the effects of SUSY, which  effectively removes otherwise quadratically 
divergent contributions to the Higgs mass from SM particles).
Therefore,  the values of this (theoretically arbitrary) parameter
$\mu$ are phenomenologically constrained to be close to $M_{\rm SUSY}$
or $M_W$ \cite{muscale}. 

The most elegant solution to the $\mu$-problem is to introduce a new singlet Higgs field $S$ and
replace $B\mu H_uH_d$ by an interaction $\sim S(H_uH_d)$. When
the extra scalar field $S$ acquires a Vacuum Expectation Value
(VEV), an effective $\mu$ term, naturally of the EW
scale, is generated automatically. 
This idea has been implemented in the Next-to-Minimal
Supersymmetric Standard Model (NMSSM) \cite{NMSSM}, described by
the Superpotential
\begin{equation}\label{WNMSSM}
W_{\rm NMSSM} = {\hat{Q}}{\hat{H}_u}{\bf{h_u}}{\hat{U}^C} 
              + {\hat{H}_d}{\hat{Q}}{\bf{h_d}}{\hat{D}^C} 
              + {\hat{H}_d}{\hat{L}}{\bf{h_e}}{\hat{E}^C}
 +\lambda\hat{S}(\hat{H}_u\hat{H}_d)+\frac{1}{3}\kappa\hat{S}^3,
\end{equation}
where $\hat{S}$ is an extra Higgs iso-singlet Superfield, 
$\lambda$ and $\kappa$ are dimensionless couplings and 
the last ($Z_3$ invariant) term  is required to explicitly break the dangerous
Peccei-Quinn (PQ) U(1) symmetry \cite{PQ}\footnote{One could also gauge the U(1)$_{\rm PQ}$ 
group, so that the $Z_3$ symmetry is embedded in the local gauge symmetry \cite{Z3}.}.  
(See Ref.~\cite{slightly} for NMSSM Higgs sector phenomenology with an exact or
slightly broken PQ symmetry.)
Furthermore, a $Z_2^R$ symmetry can be imposed
to avoid the so-called `domain-walls problem' of the NMSSM \cite{DW}, through harmless tadpoles breaking
the global $Z_3$ symmetry \cite{KT}. (Alternative formulations -- known
as the Minimal Non-minimal Supersymmetric Standard Model (MNSSM) and 
new Minimally-extended Supersymmetric Standard Model (nMSSM) -- 
vetoing the presence of the $\sim \kappa\hat{S}^3$ term and allowing instead suitable linear tadpole terms 
through the enforcement of discrete $R$-symmetries --
also exist \cite{other-non-minimal}.)
Another positive feature of all these non-minimal SUSY models is that they predict the existence of a (quasi-)stable
singlet-type neutralino (the singlino) that could be responsible for the Dark Matter (DM)
of the universe \cite{DM}.   
Finally, notice that, in these extended SUSY models, the singlet Superfield $\hat{S}$ has no SM gauge group charge
(so that MSSM gauge coupling unification is preserved)  and that one can comfortably explain the baryon asymmetry of the Universe
by means of a strong first order EW phase transition \cite{baryon1} (unlike
the MSSM, which would require a light top squark and Higgs boson barely compatible
with current experimental bounds \cite{baryon2}).  

Clearly, in eq.~(\ref{WNMSSM}), upon EWSB a 
 VEV will be generated for the real scalar 
component of $\hat S$ (the singlet Higgs
field), $<S>$, alongside those of the two doublets  $<H_u>$
and $<H_d>$ (related by the parameter $\tan\beta=
<H_u>/<H_d>$). In the absence of fine-tuning, one should expect these
three VEVs to be of the order of $M_{\rm{SUSY}}$ or $M_W$, so that now
one has an `effective $\mu$-parameter' 
\begin{equation}
\mu_{\rm{eff}}=\lambda <S>,
\end{equation}
of the required size, thus effectively solving the $\mu$-problem.

\section{\large Higgs Phenomenology in the NMSSM}

In the NMSSM, the soft SUSY-breaking Higgs sector is described by the
Lagrangian contribution
\begin{equation}
V_{\rm NMSSM}=m_{H_u}^2|H_u|^2+m_{H_d}^2|H_d|^2+m_{S}^2|S|^2
             +\left(\lambda A_\lambda S H_u H_d + \frac{1}{3}\kappa A_\kappa S^3 + {\rm h.c.}\right),
\end{equation}
where $A_\lambda$ and $A_\kappa$ are dimensionful parameters of order $M_{\rm{SUSY}}$. 

As a result of the introduction of an extra complex singlet scalar
field, which only couples to the two MSSM-type Higgs doublets, the
Higgs sector of the NMSSM comprises of a total of seven mass
eigenstates: a charged pair $h^\pm$, three CP-even Higgses
$h_{1,2,3}$ ($m_{h_1}<m_{h_2}<m_{h_3}$) and two CP-odd
Higgses $a_{1,2}$ ($m_{a_1}<m_{a_2}$). Consequently, Higgs
phenomenology in the NMSSM may be plausibly different from that of the
MSSM and extremely rich of new signals. 

For a start, the mass expressions for
the CP-even Higgses in the NMSSM can be translated into a 
modified upper bound on the lightest Higgs mass, $m_{h_1}$, as \cite{MNZ}
\begin{eqnarray}
m_{h_1}^2 \leq {\rm min}\{m_Z^2,\frac{1}{2}\kappa <S>(4\kappa
<S>+\sqrt{2}A_\kappa)\}. 
\end{eqnarray}
As the higher order corrections are similar to those in
the MSSM, the upper bound on the lightest Higgs boson
is different in the NMSSM, reaching 135--140 GeV, for maximal
stop mixing and $\tan\beta=2$ \cite{upper,Cyril} (a configuration indeed excluded
in the MSSM by LEP data). More in general, the
`little fine tuning problem', resulting in LEP failing
to detect a light CP-even Higgs boson, predicted over most
of the MSSM parameter space, is much attenuated in the NMSSM,
because the mixing among more numerous CP-even or CP-odd
Higgs fields enables light mass states being produced at LEP
yet they can remain undetected because of their reduced couplings
to $Z$ bosons \cite{Cyril}.   
  
As for future machines, chiefly the CERN Large Hadron Collider (LHC),  
quite some work has been dedicated to probing the NMSSM
Higgs sector over recent years. Primarily, there have been attempts
to extend the so-called `No-lose theorem' of the MSSM -- stating 
 that at least one MSSM Higgs boson should be observed via
the usual SM-like production and decay channels at the LHC 
 throughout the entire MSSM parameter space \cite{NoLoseMSSM} --
to the case of the NMSSM \cite{NoLoseNMSSM1}. From this perspective,
it was realised that at least one NMSSM Higgs boson should remain observable 
at the LHC over the NMSSM parameter space that does not allow any Higgs-to-Higgs 
decay. However, when the only light non-singlet (and, therefore, potentially visible) CP-even
Higgs boson, $h_1$ or $h_2$, decays mainly to two very light 
CP-odd Higgs bosons, $h\to a_1 a_1$, one
may not have a Higgs signal of statistical significance at the LHC \cite{dirk}. In fact, 
further violations to the theorem may well occur if
one enables Higgs-to-SUSY decays (e.g., into neutralino pairs, yielding invisible Higgs signals).

While the jury is still out on whether a `No-lose theorem' can be proved for the NMSSM,
we are here concerned with an orthogonal approach. We asked ourselves if a, so to say,
`More-to-gain theorem' can be formulated in the NMSSM. That is, whether there exist regions
of the NMSSM parameter space where  more Higgs
states of the NMSSM are visible at the LHC than those available within the MSSM. 
In our attempt to overview all such possibilities, we start by considering here the case
of the di-photon decay channel of a neutral Higgs boson. This mode can be successfully
probed in the MSSM, but limitedly to the case of one Higgs boson only, which is
CP-even and rather light. We will show that in the NMSSM one can instead potentially have  
up to three di-photon signals of Higgs bosons, involving not only CP-even but also CP-odd states,
the latter with masses up to 600 GeV or so. In fact, even when only one di-photon signal can
be extracted in the NMSSM, this may well be other than the $h_1$ state. When only the latter is
visible, finally, it can happen that its mass is larger than the maximum value achievable within the MSSM.
In all such cases then, the existence of a non-minimal SUSY Higgs sector would be manifest.

\section{\large Parameter Space Scan}

The choice of parameter space largely depends on the version of NMSSM
under consideration, or more specifically, on the implication of unification of parameters at some very high scale. This, in turn, leans on the technique adopted for
the breaking of SUSY, since it still remains undetected in nature. 
The advantage of assuming unification of masses and couplings in a SUSY model is twofold. Firstly, it enormously 
reduces the 
labour from a phenomenological perspective by minimising the number of parameters required to extrapolate physical information from the model.
Secondly, and theoretically more crucially, it caters to the fundamental objective of constructing a Grand Unification Theory (GUT), which was, to
a great extent, responsible for devising SUSY in the first place.  

A particular case of a  low energy NMSSM with gravity mediated SUSY breaking
was studied in \cite{cNMSSM}. As usual, in such a model, SUSY is assumed to be broken in
some hidden sector and then mediated to the physical sector
through gravitational interactions. Such a model implies
unification of the couplings and soft masses at the GUT scale,
which are then run down, using renormalisation group equations, to the weak scale 
or to some other scale at which the theory is being tested. This results in only a handful of parameters to deal with. 

For a more general study of the NMSSM Higgs sector with a wider
range of parameters, we used here the publicly available fortran code
NMHDECAY (version 1.1) developed in Ref.~\cite{NMHDECAY}. This program
computes the masses, couplings and decay widths of all the Higgs
bosons of the NMSSM in terms of its parameters at the EW
scale. For our purpose, instead of postulating unification, and without taking into account the 
SUSY breaking mechanism, we fixed the soft SUSY breaking terms to a very high value, 
so that they have little or no contribution to the outputs of the parameter scans. 
Consequently, we are left with six free parameters.

Our  parameter space includes the Yukawa couplings $\lambda$ and
$\kappa$, the soft trilinear terms $A_\lambda$ and $A_\kappa$, plus 
tan$\beta$ and $\mu_{\rm eff} = \lambda\langle S\rangle$. The
computation of the spectrum includes leading two loop terms,
EW corrections and propagator corrections. The decay
widths, however, do not include three body decays. The NMHDECAY program
also takes into account theoretical as well as
experimental constraints from negative Higgs searches at LEP,
along with the unconventional channels relevant for the NMSSM.

We have used the NMHDECAY code to scan over the NMSSM
parameter space defined through the aforementioned six parameters 
taken in the following intervals:
\begin{center}
$\lambda$ : 0.0001 -- 0.75,\phantom{aa} $\kappa$ : $-$0.65 --
+0.65,\phantom{aa} $\tan\beta$ : 1.6 -- 54,\\ $\mu$, $A_{\lambda}$,
$A_{\kappa}$ :  $-$1000 -- +1000 GeV.\\
\end{center}
Remaining soft terms which are fixed in the scan include:\\
$\bullet\phantom{a}m_{Q_3} = m_{U_3} = m_{D_3} = m_{L_3} = m_{E_3} = 2$ TeV, \\
$\bullet\phantom{a}A_{U_3} = A_{D_3} = A_{E_3} = 1.5$ TeV,\\
$\bullet\phantom{a}m_Q = m_U = m_D = m_L = m_E = 2$ TeV,\\
$\bullet\phantom{a} M_1 = M_2 = M_3 = 3$ TeV.\\

In line with the assumptions made in \cite{NoLoseNMSSM2}, the allowed decay 
modes for neutral NMSSM Higgs bosons are\footnote{Here, we use the label
$h(a)$ to signify any of the neutral CP-even(odd) Higgs bosons of the NMSSM.}:
\begin{eqnarray}
h,a\rightarrow gg,\phantom{aaa} h,a\rightarrow \mu^+\mu^-,
&&h,a\rightarrow\tau^+\tau^-,\phantom{aaa}h,a\rightarrow
b\bar b,\phantom{aaa}h,a\rightarrow t\bar t,\nn \\ h,a\rightarrow
s\bar s,\phantom{aaa}h,a\rightarrow
c\bar c,&&h\rightarrow W^+W^-,\phantom{aaa}h\rightarrow ZZ,\nn \\
h,a\rightarrow\gamma\gamma,\phantom{aaa}h,a\rightarrow
Z\gamma,&&h,a\rightarrow {\rm Higgses},\phantom{aaa}h,a\rightarrow
{\rm sparticles}. \nn
\end{eqnarray}
(Notice that for the pseudoscalar Higgses, the decay into vector
boson pairs is not allowed due to CP-conservation.) Here, 
`Higgses' refers to any final state
involving all possible combination of two Higgs bosons (neutral and/or charged)
or of one Higgs boson and a gauge vector.

The aforementioned  range of parameters is borrowed from 
Ref.~\cite{NoLoseNMSSM2}, where the authors singled out the most difficult scenarios for
NMSSM Higgs discovery at the LHC (using the mentioned code). 
In this paper they also concluded, as a follow up of
the NMSSM `No-Lose theorem', that the region where no Higgs is observable, due
to its decays into the unconventional channel $h\to a_1 a_1$,
comprises of only about 1\% of the entire parameter space of the
NMSSM. In short, our idea is to explore the same range of parameter space
and look for the `best case scenarios' where the discovery of one
or more Higgses at the LHC is possible above and beyond what is
predicted in the MSSM.

We have performed our scan over several millions of randomly
selected points in the specified parameter space. The output, as
stated earlier, contains masses, Branching Ratios (BRs) and couplings of
the NMSSM Higgses, for all the points which are not forbidden by
the various constraints. The points which violate the constraints
are eliminated. The surviving data points are then used to determine the
cross-sections for NMSSM Higgs hadro-production  by using an adapted
version of the codes 
described in \cite{WJSZK}. As the SUSY mass scales 
have been arbitrarily set well above the EW one (see above), 
the NMSSM Higgs production modes
exploitable in simulations at the LHC are those involving couplings to
heavy ordinary matter only, i.e., (hereafter, $V=W^\pm,Z$ and $Q=b,t$)
for neutral Higgs production (where the last two channels are only allowed
for CP-even Higgs production):
\vspace*{-0.2cm}
$$gg\to {\rm{Higgs}}~({\rm{gluon-fusion,~via~heavy-quark~loops}}),$$
$$gg\to Q\bar Q~{\rm{Higgs}}~({\rm{heavy-quark~associated~production}}),$$
$$qq\to qq V^{*}V^{*}\to qq~{\rm{Higgs}}~({\rm{vector-boson-fusion}}),$$
$$q\bar q\to V~{\rm{Higgs}}~({\rm{Higgs-strahlung}}).$$
(These are the so-called `direct' Higgs production modes.) Here, 
`Higgs' refers to any possible neutral Higgs boson.

Production and decay rates for NMSSM neutral Higgs bosons 
have then been multiplied together 
to yield inclusive event rates, assuming a LHC luminosity of 
100 fb$^{-1}$ throughout.

\section{\large Spectrum Configuration and Inclusive Event Rates}

As an initial step towards the analysis of the data, we computed the total cross-section
times BR into $\gamma\gamma$ pairs against each of
the six parameters of the NMSSM, for each Higgs boson.
We have assumed all production modes described in the above 
Section and started by computing total (i.e., fully inclusive) 
rates\footnote{After verifying that the bulk of the signal rates is due to gluon-gluon
fusion (even at large Higgs masses), we have eventually decided -- for simplicity -- to limit ourselves to 
emulate only this channel. Hence, all the results found below suffer from a slight under-estimate
of the signal rates.}. We are here  focusing on the $\gamma\gamma$ decay mode since
it is the most promising channel for the discovery of a (neutral) 
Higgs boson at the LHC in the moderate Higgs mass range (say, below
130 GeV). In fact, since the tail
of the $\gamma\gamma$ background falls rapidly with increasing
invariant mass of the di-photon pair,
signal peaks for heavier Higgses could also be visible in addition
to (or instead of) the lightest one, although the cross-section for these
processes is relatively very low \cite{abdel,Jacobs}. As the starting point of our
numerical study, based on the
ATLAS analysis of Ref.~\cite{ATLASTDR}, we argue that cross-section times
BR rates of 10 fb or so are potentially interesting from a
 phenomenological point of view, in the sense that they may yield visible signal
events, the more so the heavier the decaying Higgs state
(also because the photon detection efficiency grows with the Higgs mass \cite{ATLASTDR}).
(See Ref.~\cite{Erice} for a preliminary account in this respect.)

Tab.~\ref{tab:1} recaps the potential observability of one or
more NMSSM Higgs states in the di-photon mode at the LHC, under the
above assumptions. It is obvious from the table that one light CP-even
Higgs should be observable almost throughout the NMSSM parameter space
(in line with the findings of Ref.~\cite{NoLoseNMSSM2}). However,
there is also a fair number of
points where two Higgses may be visible simultaneously ($h_1$ and $h_2$ or --
more rarely -- $h_1$ and $a_1$),
while production and decay of the three lightest Higgses  ($h_1$, $h_2$ and $a_1$) 
at the same time, although possible, 
occurs for only a very small number of points in the parameter space.
Furthermore, the percentage of points for which only the second lightest
Higgs state is visible is also non-negligible. These last two conditions are clearly 
specific to the NMSSM, as they are never realised in the MSSM. Furthermore, while
the lone detection of the lightest CP-even NMSSM Higgs boson may mimic a similar
signal from the corresponding state in the MSSM, the reconstructed mass may well be
beyond the upper mass limit in the MSSM, this possibility also pointing towards 
the evidence of a NMSSM Higgs sector. Finally, none of the two heaviest NMSSM neutral Higgs states
($h_3$ and $a_2$) will be visible in the di-photon channel at the LHC (given their
large masses). 

We have then plotted the cross-sections times BR for the
three potentially observable Higgses against the various NMSSM parameters.
These plots, shown in Figs.~1--6, reveal that the distribution of possibly visible points 
(i.e., of those yielding cross-section time BR rates in excess of 10 fb)
is quite homogeneous over the NMSSM parameter space and not located in some specific
parameter areas (i.e., in a sense, not `fine-tuned'). The distribution of the same points
as a function of the corresponding Higgs masses can be found in Fig.~7. Of particular relevance
is the distribution of points in which only the NMSSM $h_1$ state is visible, when its mass is beyond
the upper mass limit for the corresponding CP-even MSSM Higgs state, which is shown in Fig.~8\footnote{Notice that
the value obtained for $m_{h_1}^{\rm{max}}$  from NMHDECAY version 1.1, of 
$\sim 130$ GeV, based on the leading two-loop approximations
described in \cite{NMHDECAY}, is a few GeV lower than the value declared in Sect.~1.
Besides, for consistency, we use the value of 120 GeV (obtained at the same level of accuracy) as
upper mass limit on the lightest CP-even Higgs boson of the MSSM.
(Notice that a slightly modified  $m_{h_1}^{\rm{max}}$  value is obtained for the NMSSM from 
NMHDECAY version 2.1 \cite{NMHDECAY2},  
because of the improved mass 
approximations with respect to the earlier version of the program adopted here.)
Eventually, when the LHC is on line,
the exercise that we are proposing can be performed with the then state-of-the-art calculations.}.
This plot reveals that about
93\% of the NMSSM $h_1$ masses visible alone are expected to be within 2--3 GeV beyond the MSSM bound,
hence the two models would be indistinguishable\footnote{Other than an experimental di-photon 
mass resolution  of 2 GeV or so \cite{ATLASTDR}
one should also bear in mind here that the mass bounds in both models come with a
theoretical error of comparable size.}. Nonetheless, there is a fraction of a percent of such points
with $m_{h_1}$ values even beyond 125 GeV or so (the higher the mass the smaller the density, though), which should indeed allow one to distinguish
between the two models. Moreover, 
by studying the cross-section times BR of the Higgses when two of them
are observable against their respective mass differences, Figs.~9--11,
one sees that the former are 
larger than the typical mass resolution in the di-photon channel, 
so that the two decaying objects should
indeed appear in the data as separate resonances. (We have also verified, though not shown here,
that their decay widths are small compared to the detector resolution in $M_{\gamma\gamma}$.) 

Next, we have proceeded to a dedicated parton level analysis of signal and
background processes, the latter involving both tree-level $q\bar q\to \gamma\gamma$ and 
one-loop $gg \to \gamma\gamma$ contributions. We have adopted standard cuts on the two 
photons \cite{ATLASTDR}:
$ p_T^\gamma>  25$ GeV and
$|\eta^\gamma| <$  2.4 on transverse energy and pseudorapidity, respectively. 
As illustrative examples of a possible NMSSM Higgs phenomenology appearing at the LHC
in the di-photon channel, we have picked up the following three configurations:
\begin{enumerate}
\item  $\lambda=$ 0.6554, $\kappa=$ 0.2672, $\mu=-426.48$ GeV, $\tan\beta=$ 2.68, $A_{\lambda}=-963.30$ GeV, $A_{\kappa}= 30.48$ GeV; 
\item  $\lambda=$ 0.6445, $\kappa=$ 0.2714, $\mu=-167.82$ GeV, $\tan\beta=$ 2.62, $A_{\lambda}=-391.16$ GeV, $A_{\kappa}= 50.02$ GeV;
\item  $\lambda=$ 0.4865, $\kappa=$ 0.3516, $\mu= 355.63$ GeV, $\tan\beta=$ 2.35, $A_{\lambda}= 519.72$ GeV, $A_{\kappa}=-445.71$ GeV.
\end{enumerate}
The first is representative of the case in which only the NMSSM $h_1$ boson is visible, but with 
mass larger than the MSSM upper limit on the corresponding Higgs state. The second and third
refer instead to the case when also the $h_2$ or $a_1$ state are visible, respectively.
The final results are found in Fig.~12. The corresponding mass resonances are clearly
visible above the continuum di-photon background and discoverable beyond the $5\sigma$
level. Indeed, similar situations can be found for each of the combinations listed in 
Tab.~\ref{tab:1} and most of these correspond to phenomenological scenarios which are 
distinctive of the NMSSM and that cannot be reproduced in the MSSM. 

\section{Conclusion}

In summary, we have shown that there exists the possibility of establishing
a `More-to-gain theorem' within the NMSSM, as compared to what is 
expected in the MSSM, in terms of novel Higgs signals appearing in the di-photon
discovery channel which can be ascribed to the former but not to the latter. 
We have showed this to be the case for a few selected NMSSM 
parameter points, by performing a proper signal-to-background analysis at the partonic
level. However, a similar numerical study can easily be extended to encompass 
 sizable regions of the NMSSM parameter space. While the bulk of the latter is in a 
configuration degenerate with the MSSM case (as far as di-photon signals at the LHC are
concerned), non-negligible areas exist where further phenomenological studies have the 
potential to unveil a non-minimal nature of the underlying SUSY model.

To this end, NMSSM benchmark scenarios, amenable to experimental investigation in the context
of full Monte Carlo (MC) analyses, based on event generation and detector simulation,
are currently being devised \cite{Alliance}. Also, un upgraded version of HERWIG \cite{HERWIG},
suitable for MC event generation in the context of non-minimal SUSY models, is currently being prepared
\cite{preparation}.
 
\section*{Acknowledgments}

We acknowledge useful email exchanges with Cyril Hugonie. 
SM would like to thank the British Council under the
`Alliance: Franco British Partnership Programme 2004 (Project Number: PN 04.051)' for travel support
in connection with this research. He is also grateful to Fawzi Boudjema and Genevieve Belanger
for their kind hospitality at LAPP-TH.

\clearpage


\begin{table}
\centering\begin{tabular}{|c|ll|c|}
\hline Higgs Flavor & \multicolumn{2}{c|}{Points Visible} & Percentage\\ \hline
\hline \multirow{5}{*}{$h_1$} & Total: & 1345884 & 99.7468 \\
 & Alone: & 1345199 & 99.6961 \\
 & With $h_2$: & 528 & 0.0391 \\
 & With $a_1$: & 152 & 0.0113 \\
 & With $h_2$ and $a_1$: & 5 & 0.0004 \\ \hline
\multirow{5}{*}{$h_2$} & Total: & 1253 & 0.0929 \\
 & Alone: & 717 & 0.0531 \\
 & With $h_1$: & 528 & 0.0391 \\
 & With $a_1$: & 3 & 0.0002 \\
 & With $a_1$ and $a_1$: & 5 & 0.0004 \\ \hline
$h_3$ & Total: & 0 & 0 \\ \hline
\multirow{5}{*}{$a_1$} & Total: & 165 & 0.0122 \\
 & Alone: & 5 & 0.0004 \\
 & With $h_1$: & 152 & 0.0113 \\
 & With $h_2$: & 3 & 0.0002 \\
 & With $h_1$ and $h_2$: & 5 & 0.0004 \\ \hline
$a_2$ & Total: & 0 & 0 \\ \hline
\end{tabular}
\caption{\label{tab:1} Higgs events potentially visible at the LHC through the $\gamma\gamma$ decay mode
(i.e., those yielding cross-section time BR rates of order 10 fb or upwards). Percentage refers to the
portion of NMSSM parameter space involved for each discovery scenario. }
\end{table}

\clearpage

\begin{figure}
\begin{tabular}{ccc}
\hspace*{-1.5truecm}\includegraphics[scale=0.35]{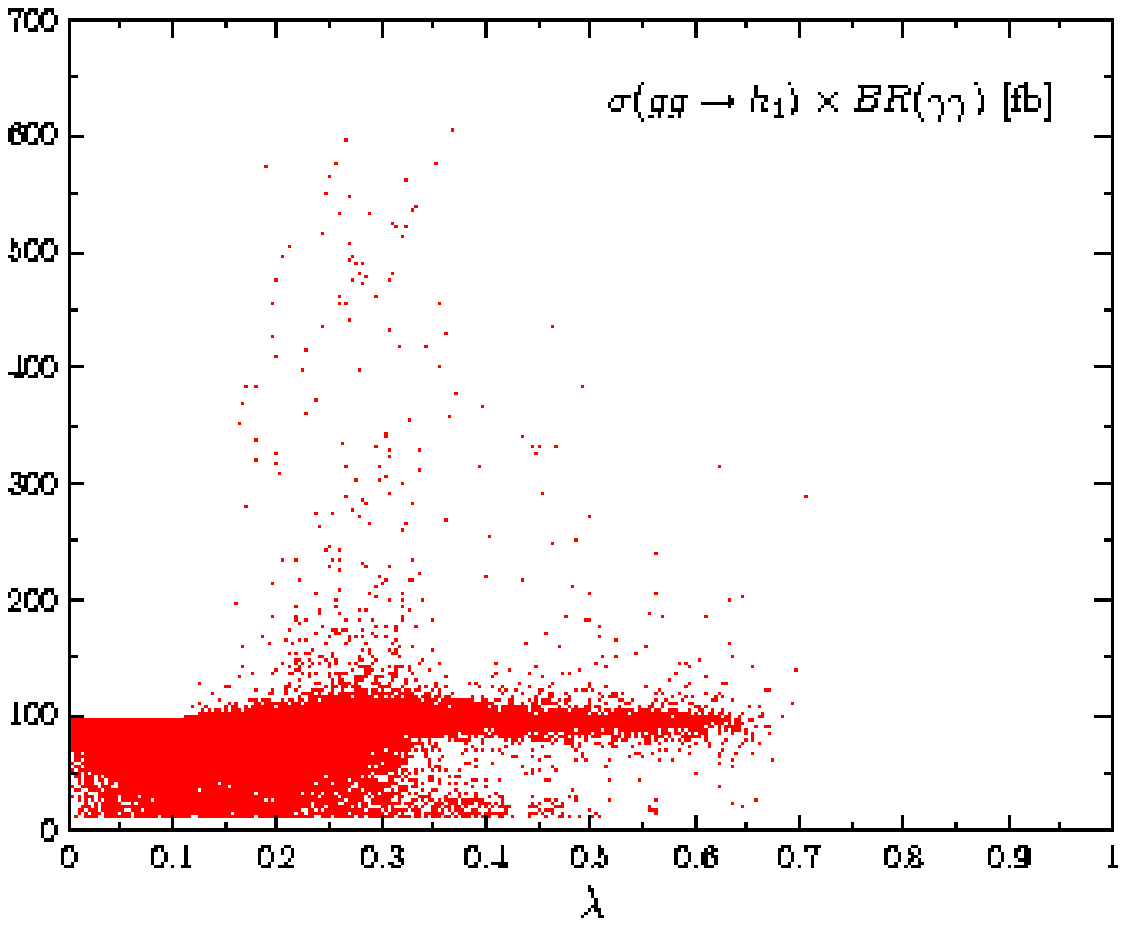}&\hspace*{-1.5truecm}\includegraphics[scale=0.35]{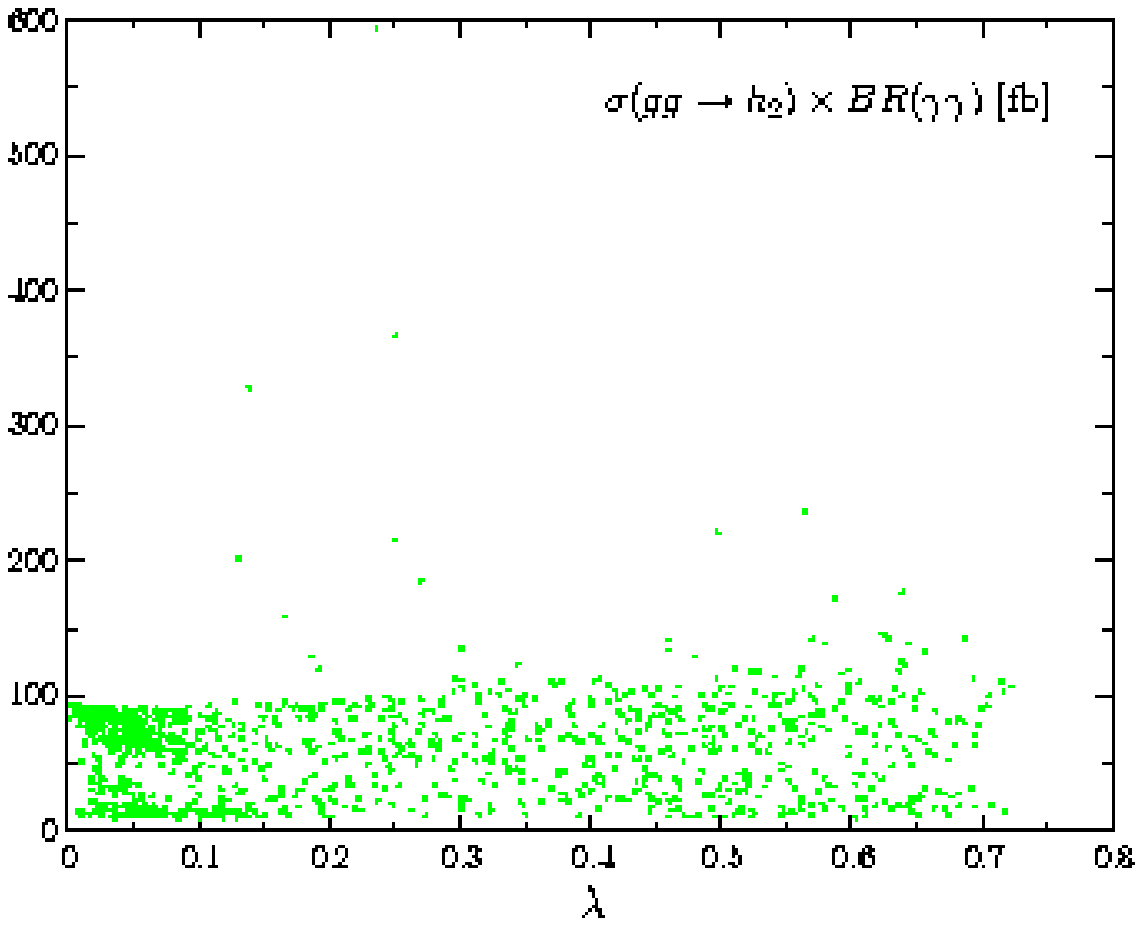} &\hspace*{-1.5truecm}\includegraphics[scale=0.35]{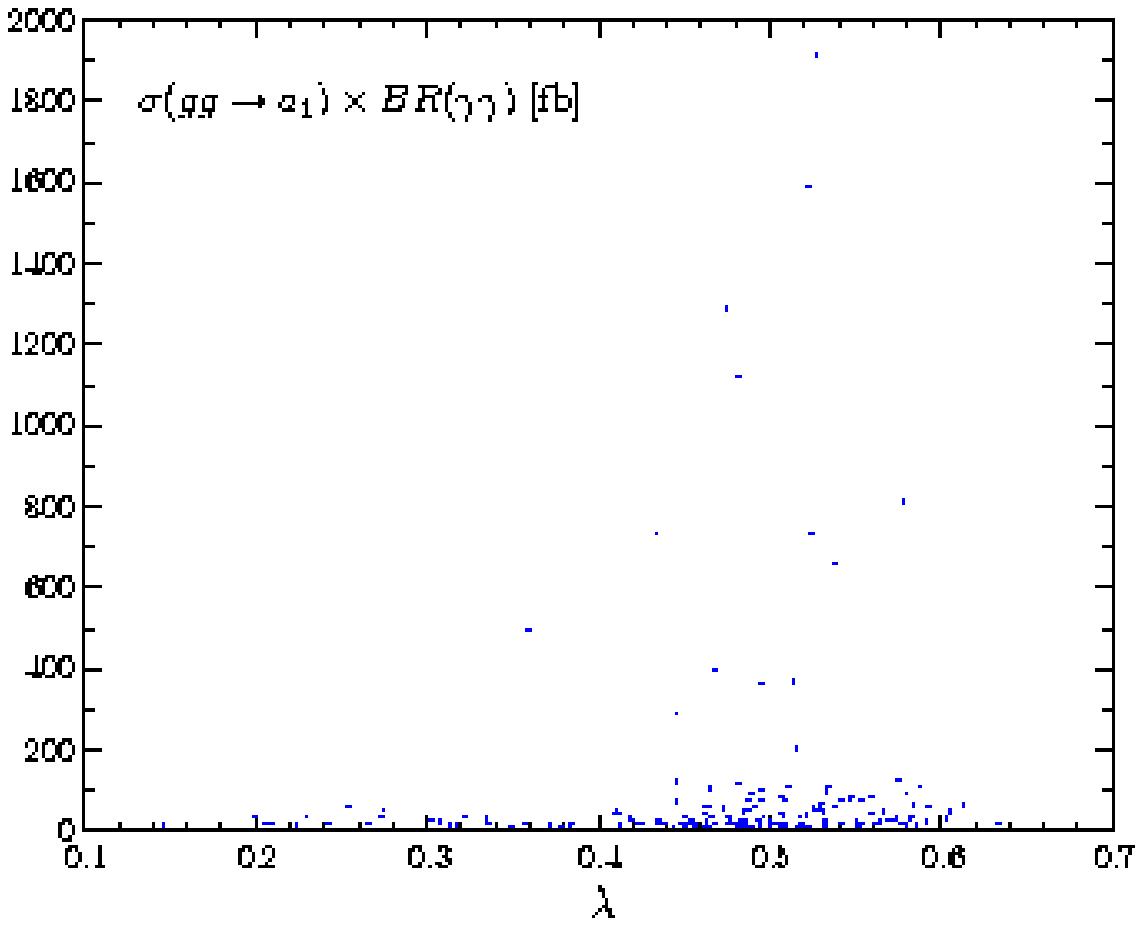}\\
\hspace*{-1.5truecm}\includegraphics[scale=0.35]{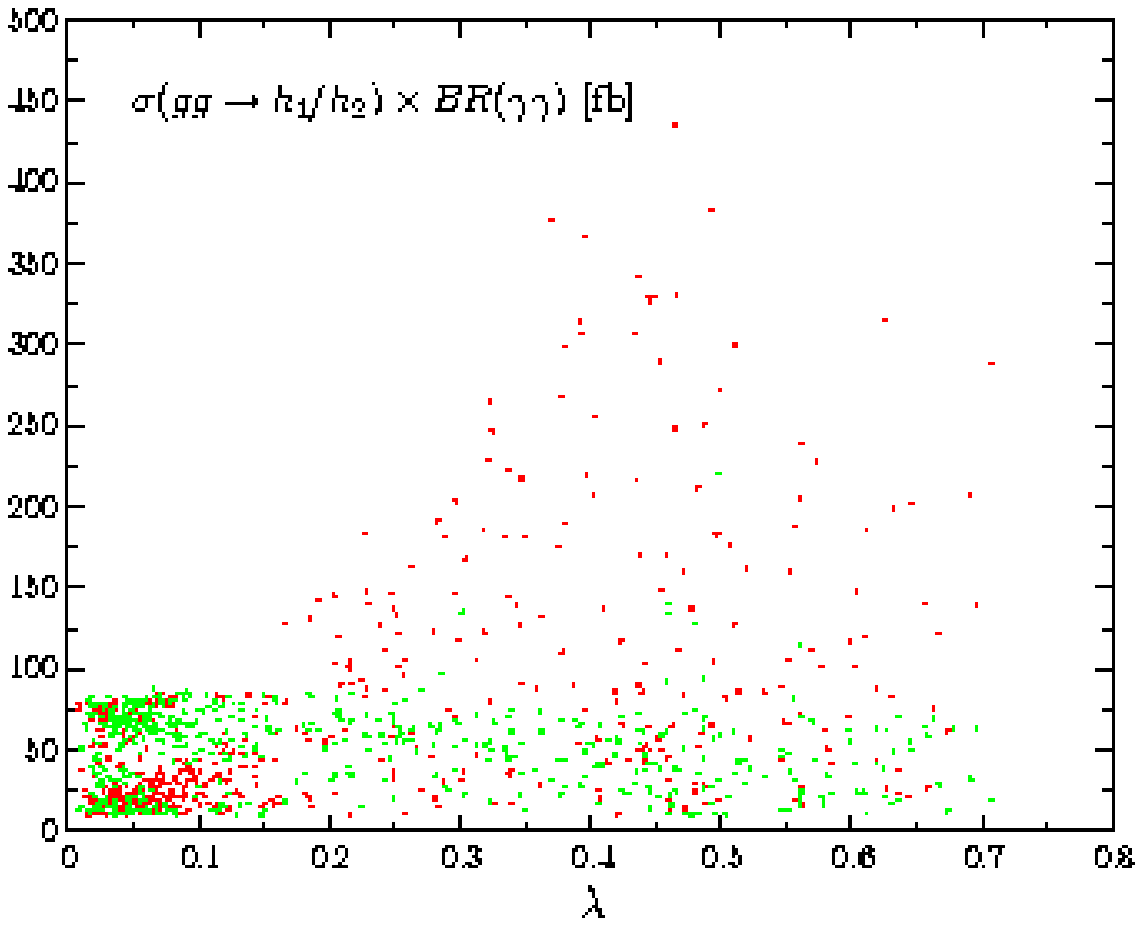}&\hspace*{-1.5truecm}\includegraphics[scale=0.35]{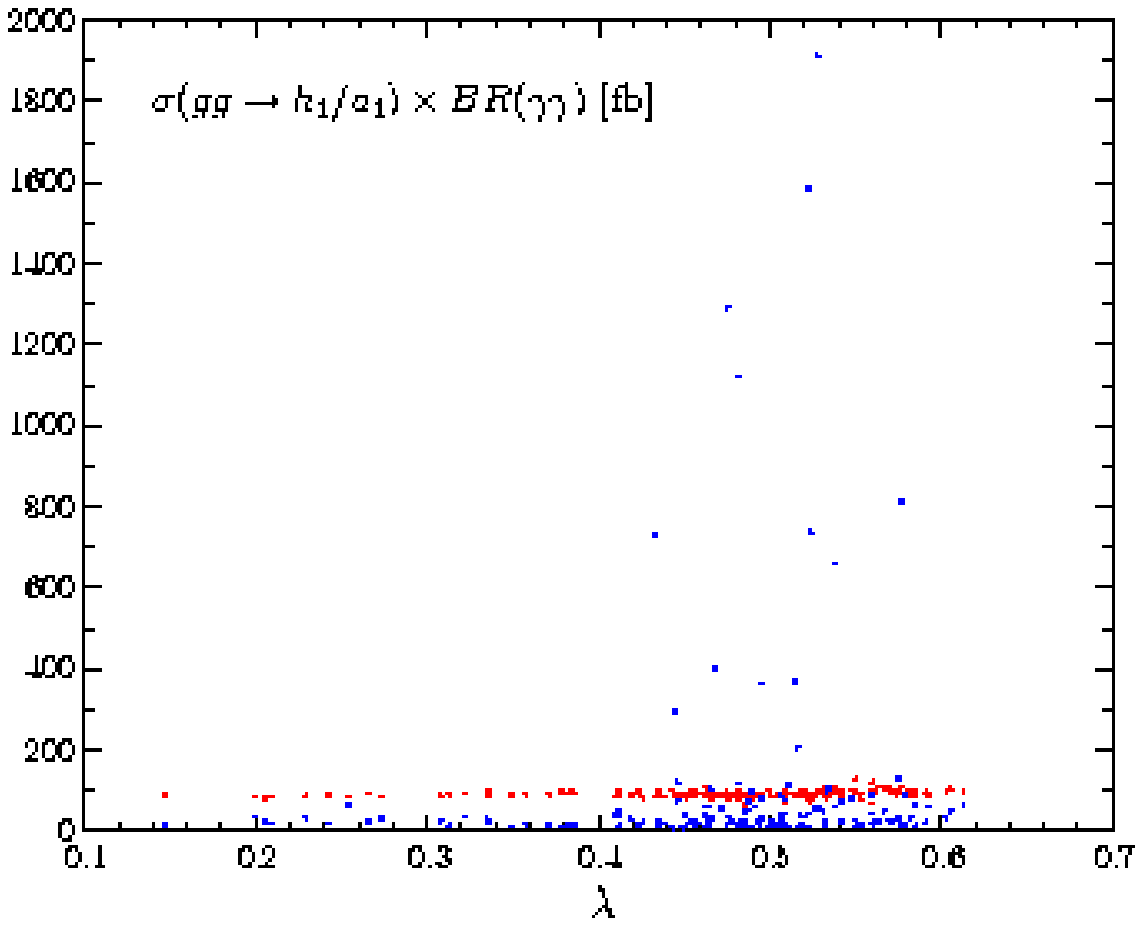}&\hspace*{-1.5truecm}\includegraphics[scale=0.35]{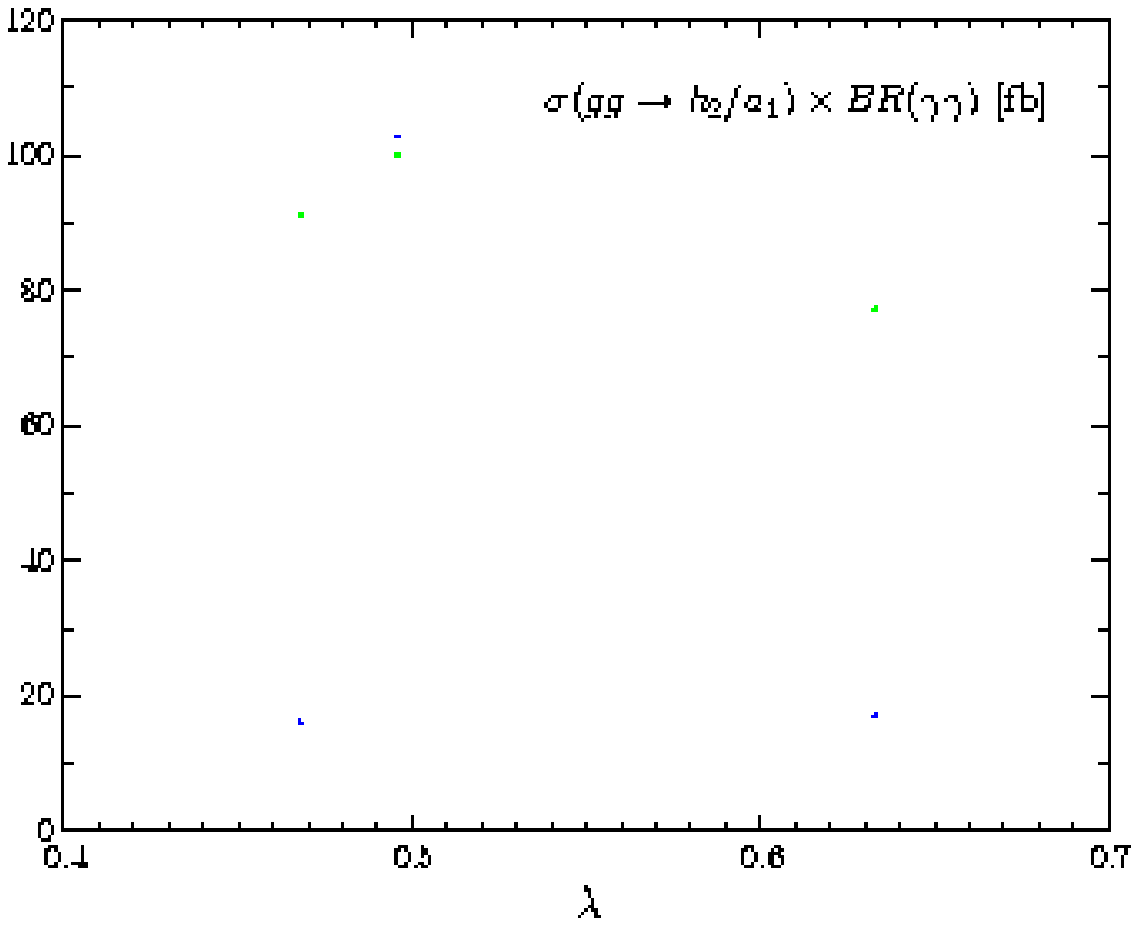}
\end{tabular}
\caption{Cross-section times BR of $h_1$ (red), $h_2$ (green) and $a_1$ (blue), when potentially visible individually and when two of these are potentially visible simultaneously, plotted against the parameter $\lambda$.}
\end{figure}

\begin{figure}
\begin{tabular}{ccc}
\hspace*{-1.5truecm}\includegraphics[scale=0.35]{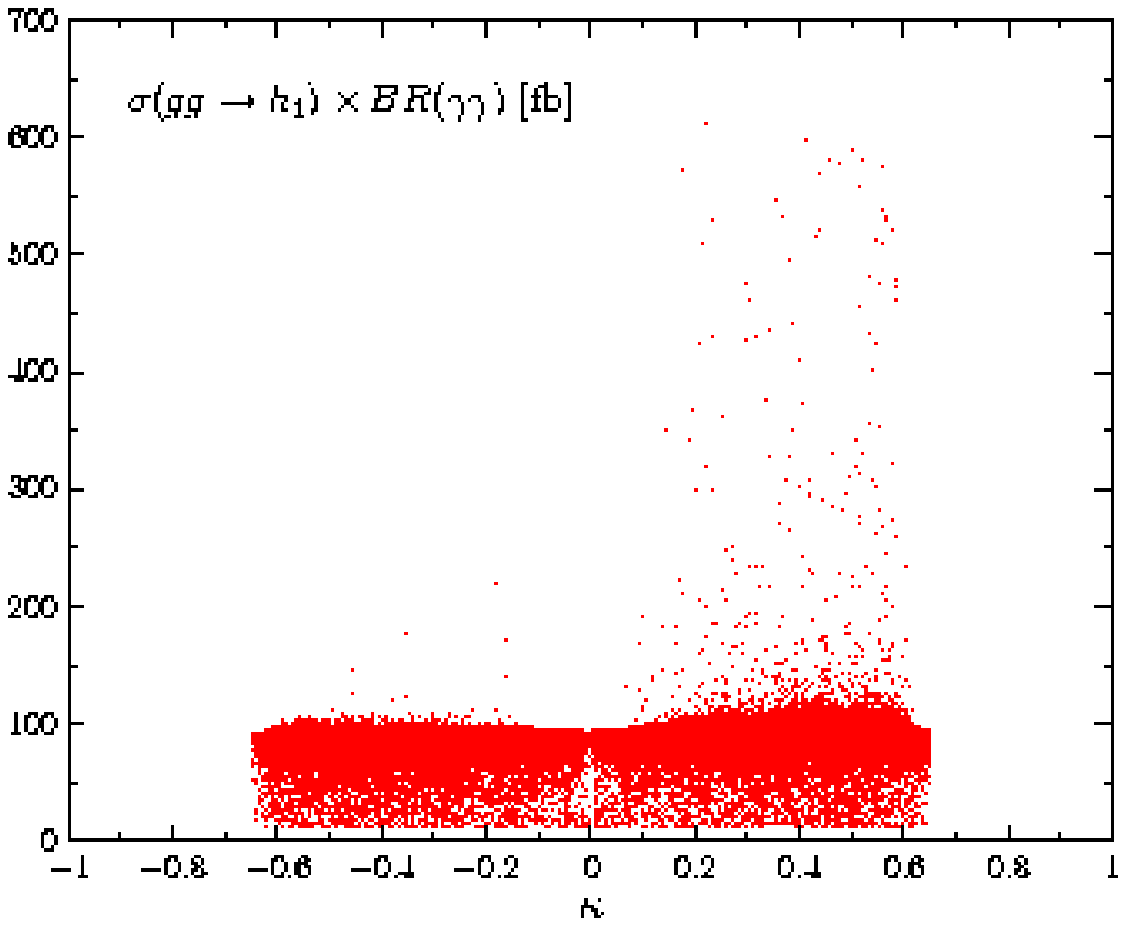}&\hspace*{-1.5truecm}\includegraphics[scale=0.35]{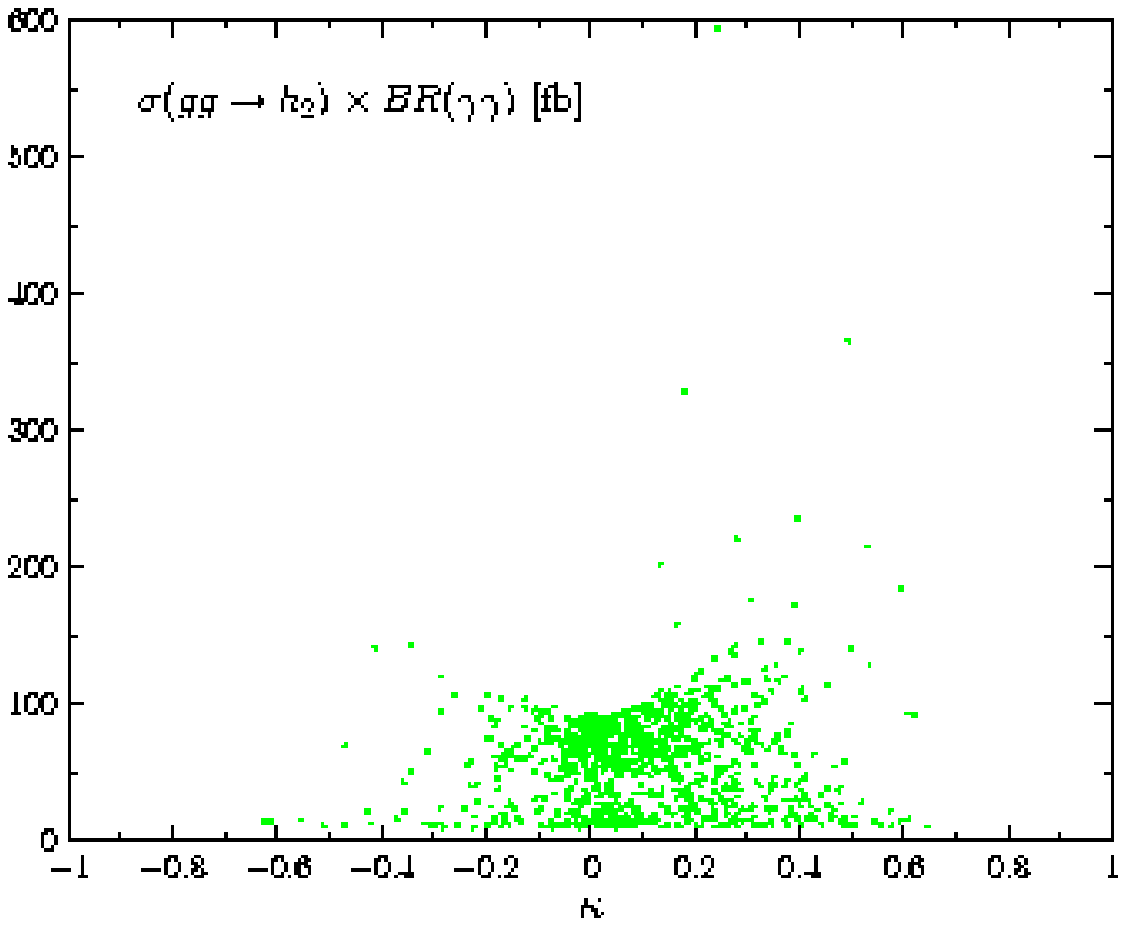} &\hspace*{-1.5truecm}\includegraphics[scale=0.35]{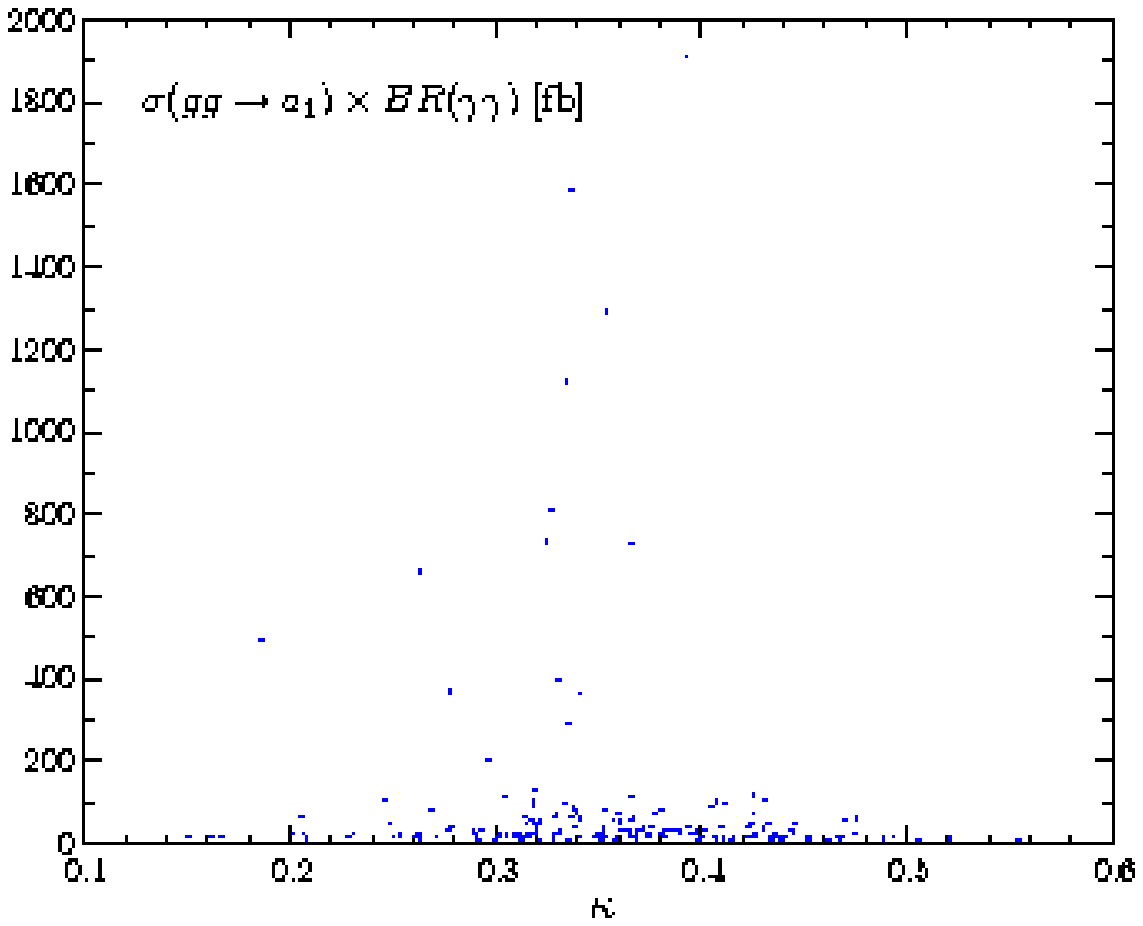}\\
\hspace*{-1.5truecm}\includegraphics[scale=0.35]{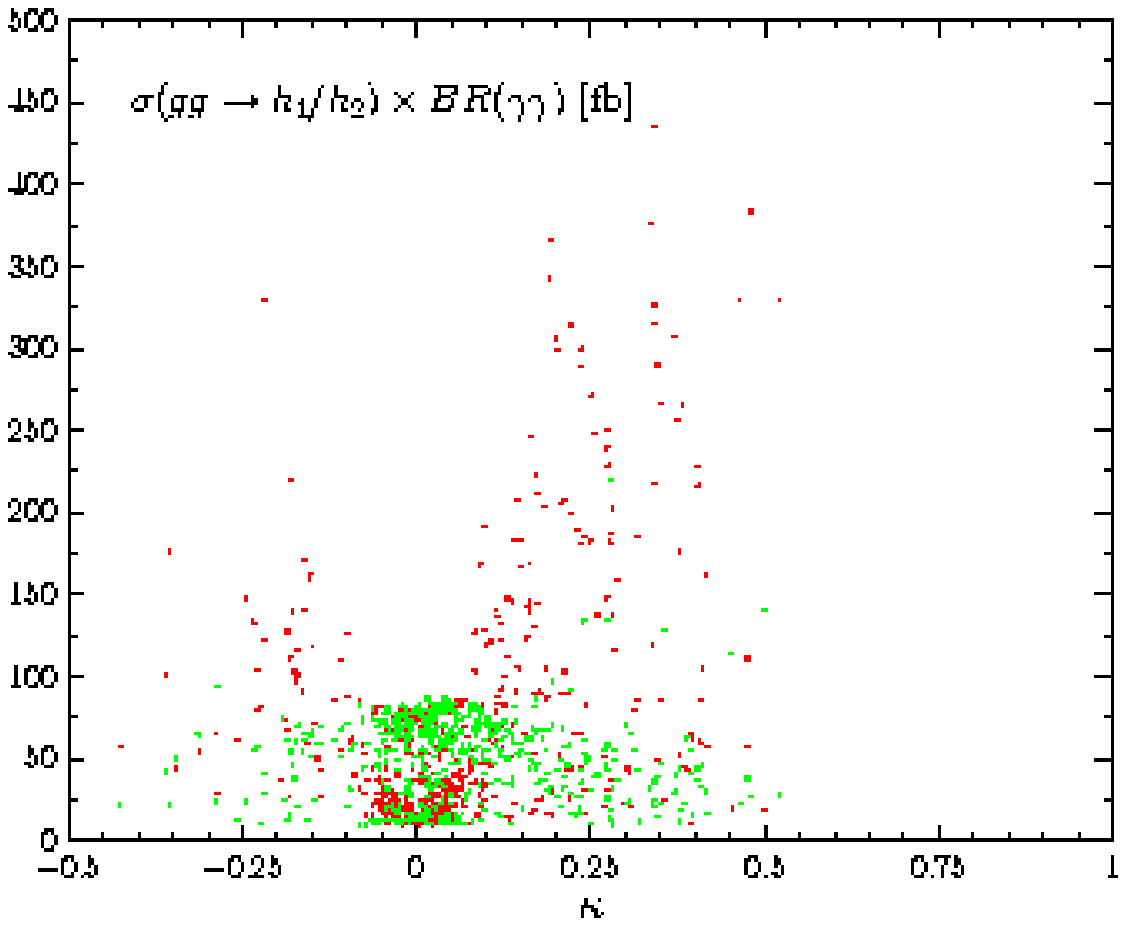}&\hspace*{-1.5truecm}\includegraphics[scale=0.35]{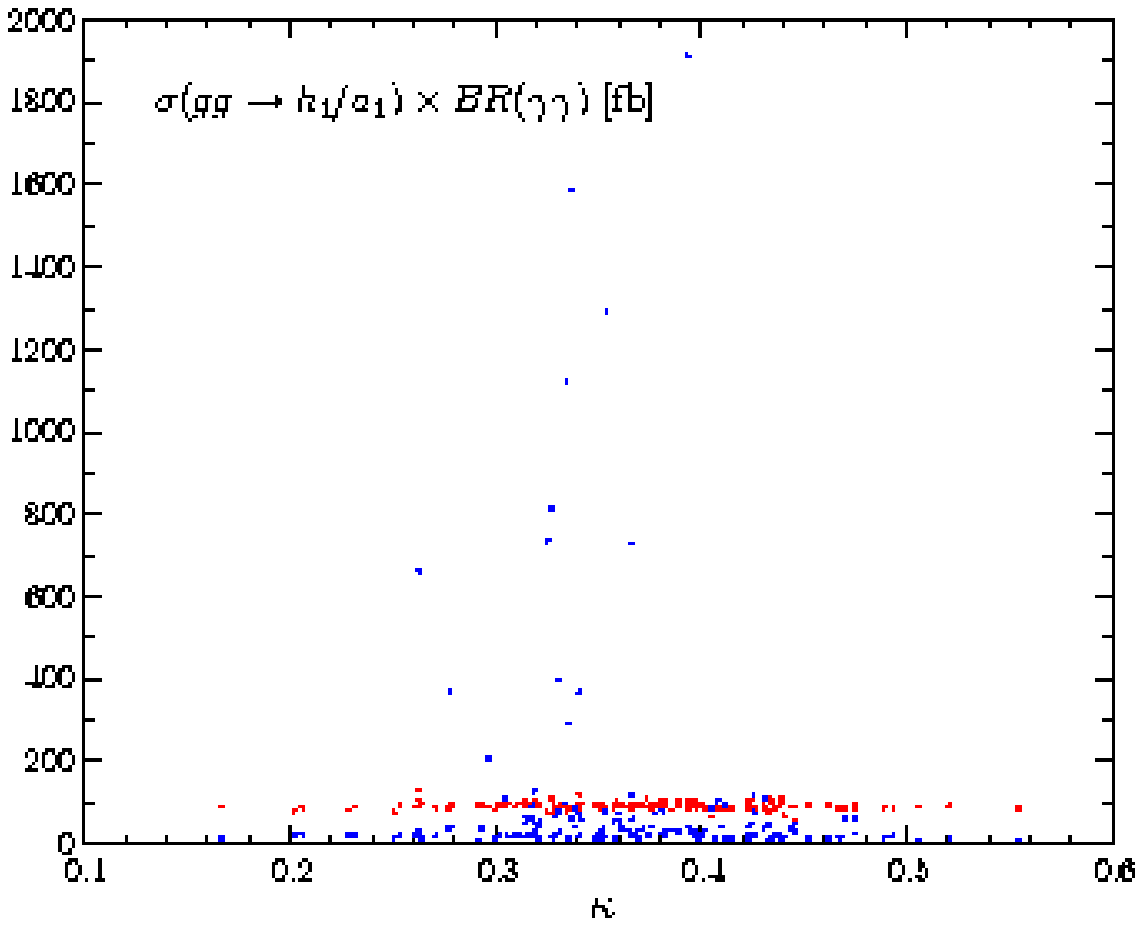}&\hspace*{-1.5truecm}\includegraphics[scale=0.35]{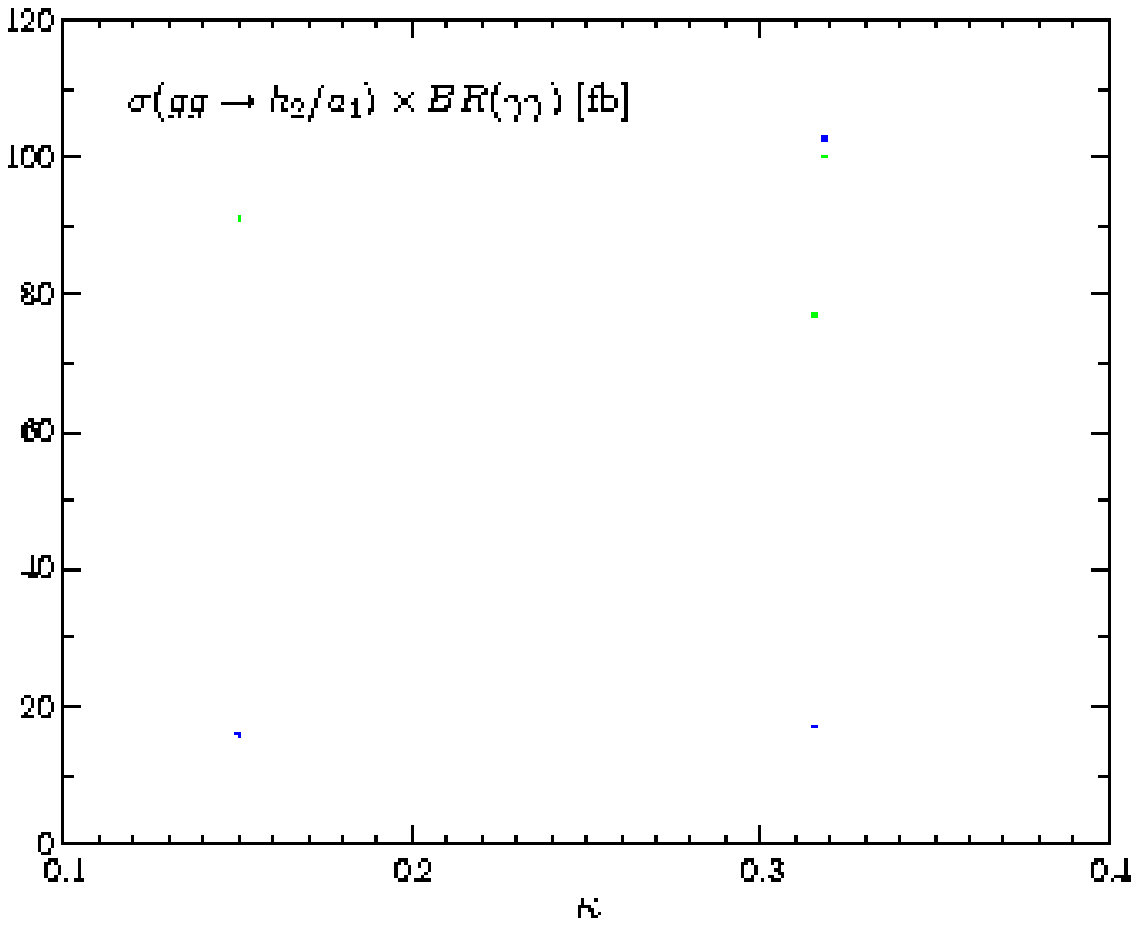}
\end{tabular}
\caption{Cross-section times BR of $h_1$ (red), $h_2$ (green) and $a_1$ (blue), when potentially visible individually and when two of these are potentially visible simultaneously, plotted against the parameter $\kappa$.}
\end{figure}

\clearpage

\begin{figure}
\begin{tabular}{ccc}
\hspace*{-1.5truecm}\includegraphics[scale=0.35]{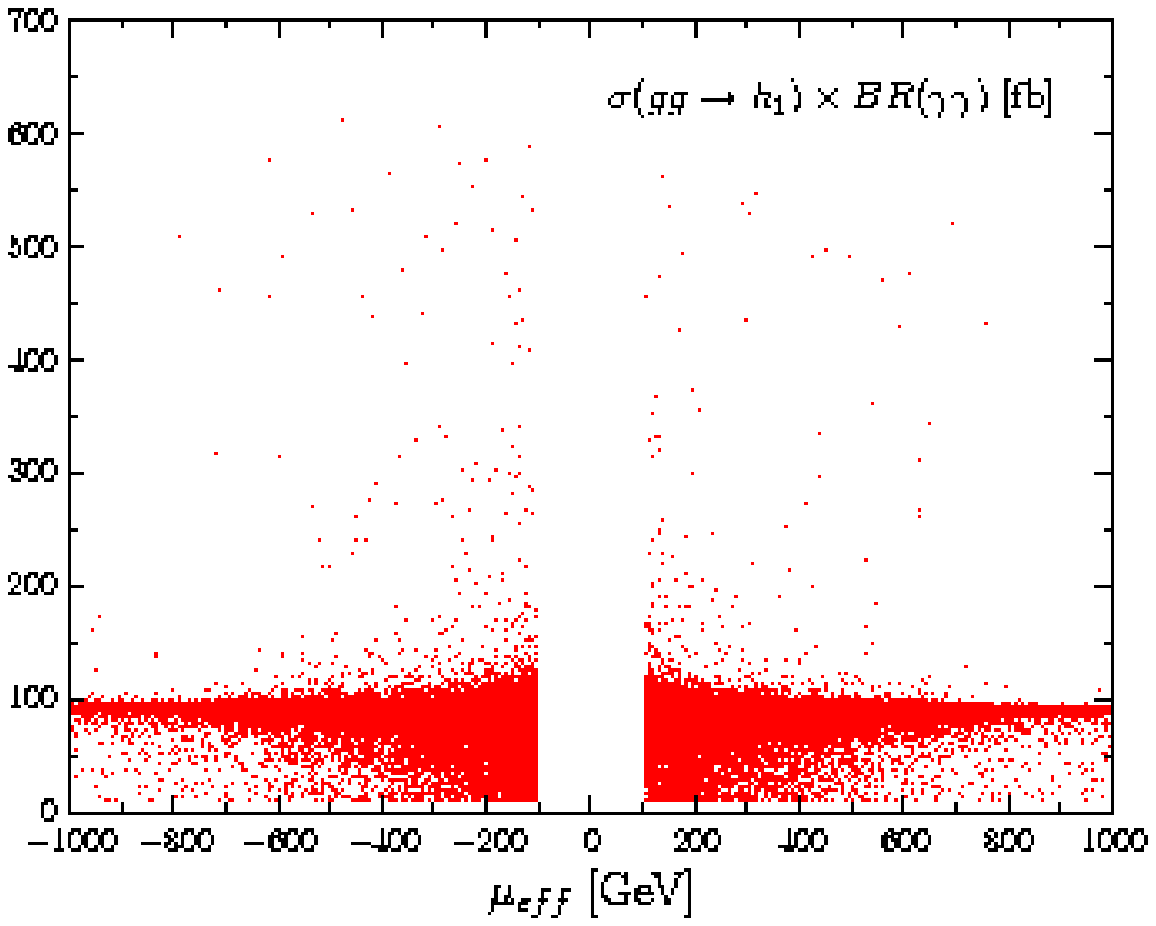}&\hspace*{-1.5truecm}\includegraphics[scale=0.35]{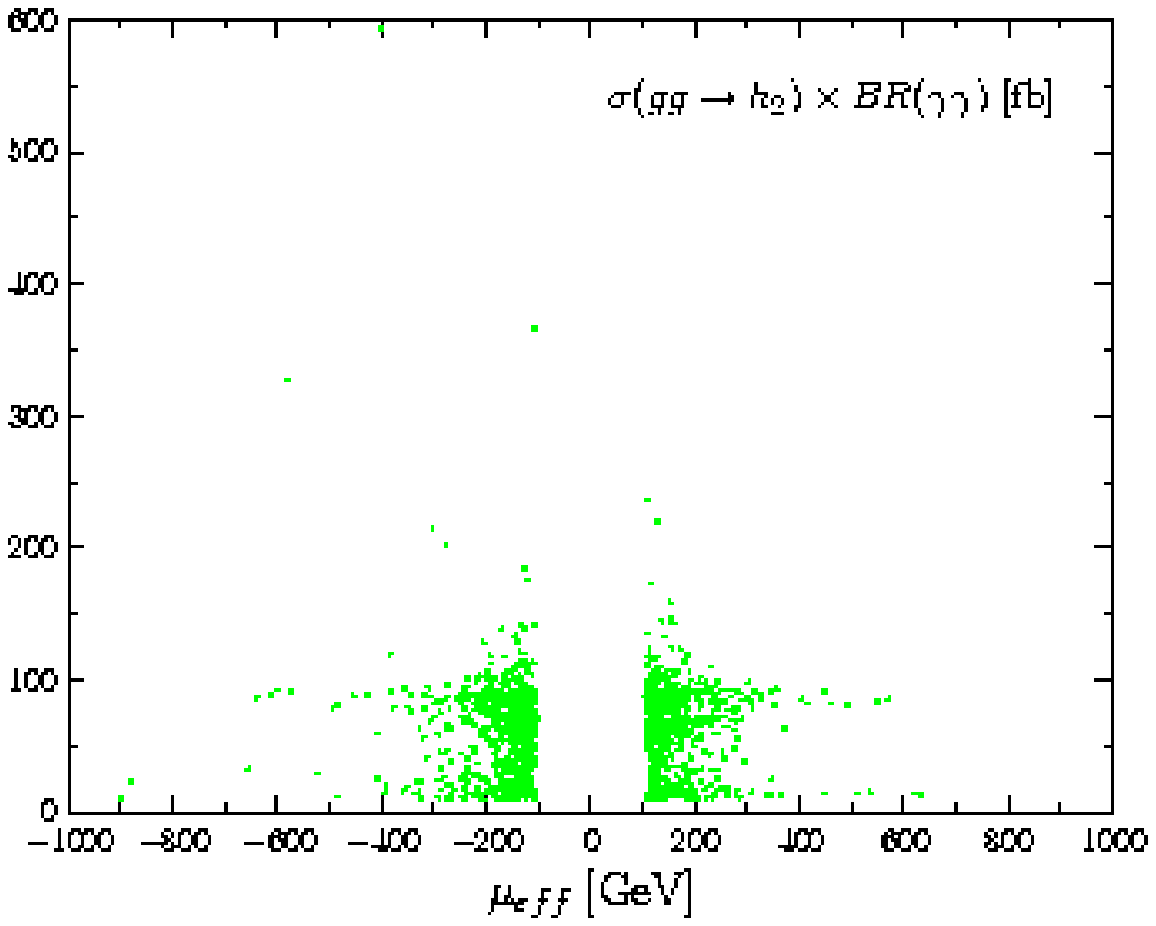} &\hspace*{-1.5truecm}\includegraphics[scale=0.35]{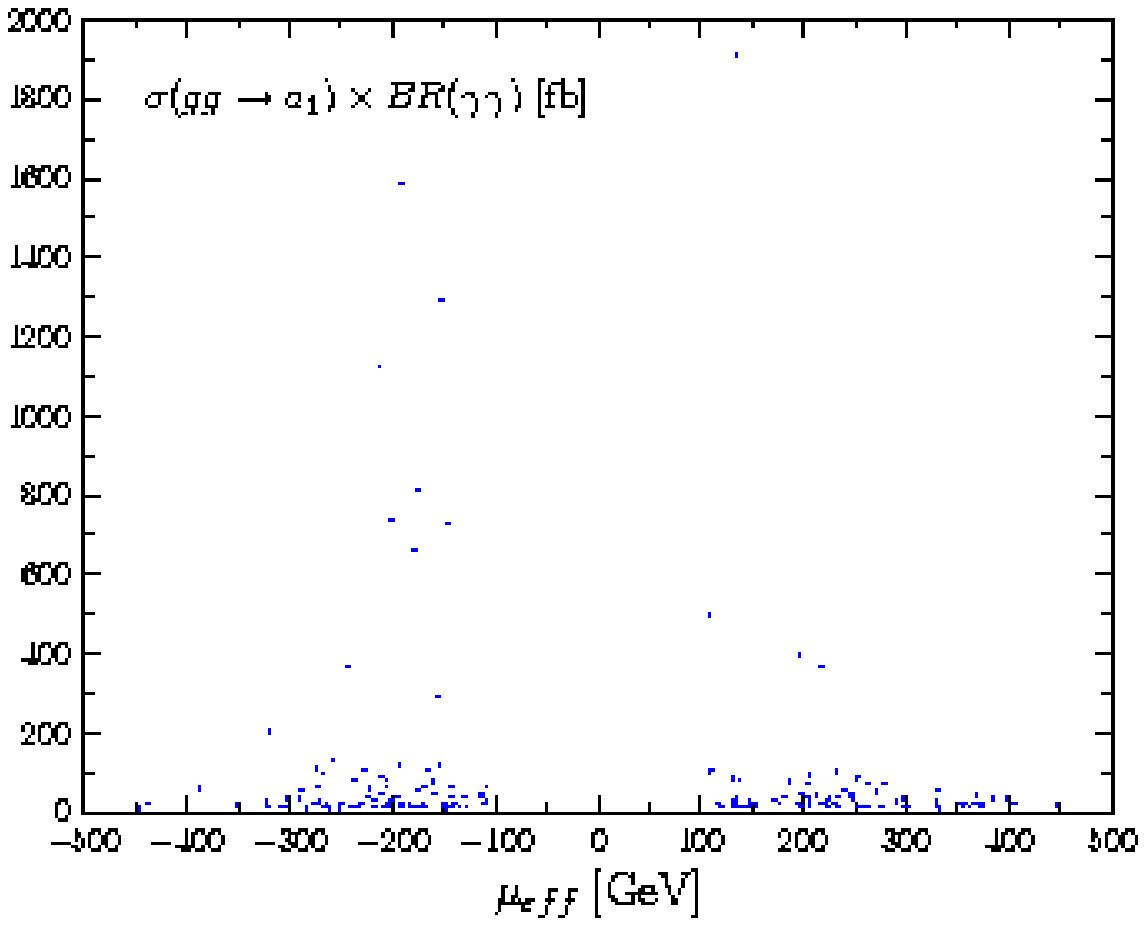}\\
\hspace*{-1.5truecm}\includegraphics[scale=0.35]{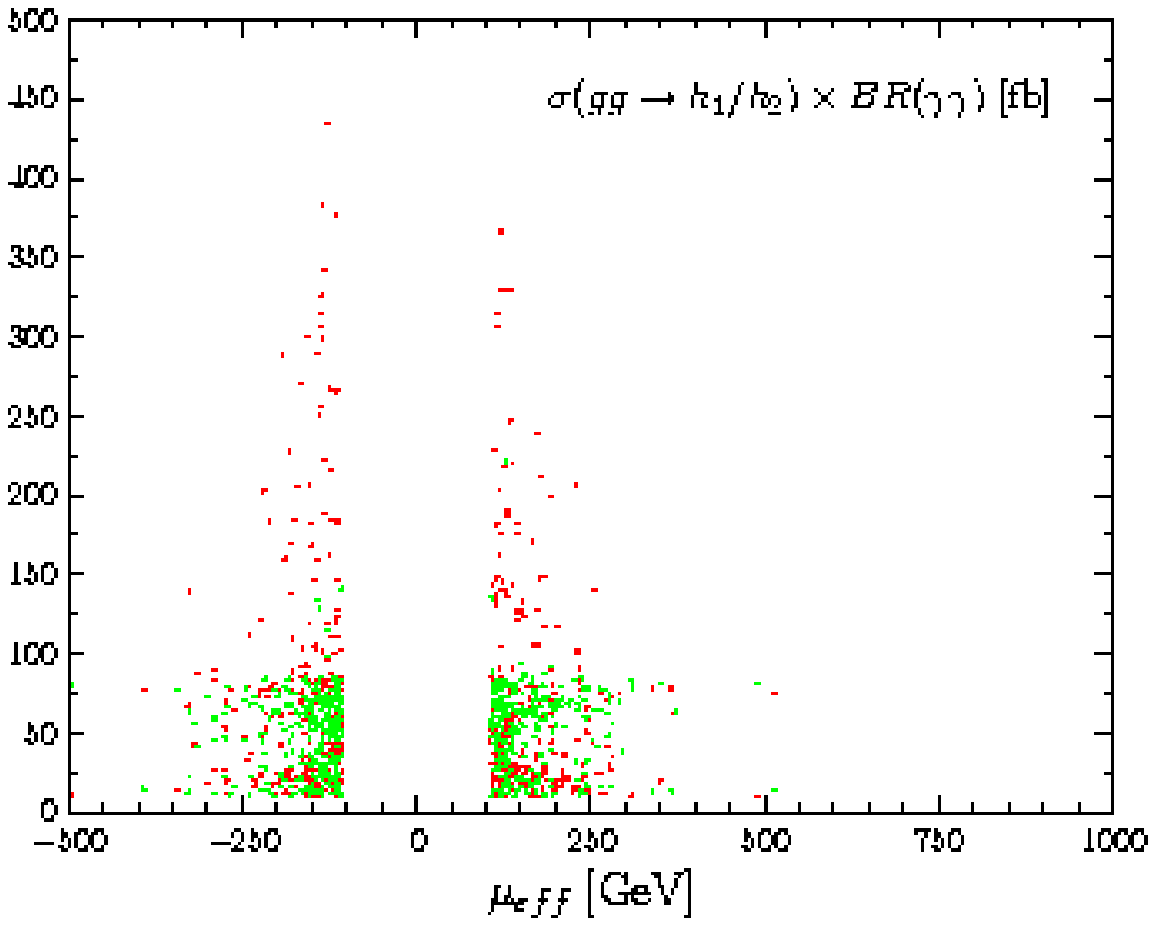}&\hspace*{-1.5truecm}\includegraphics[scale=0.35]{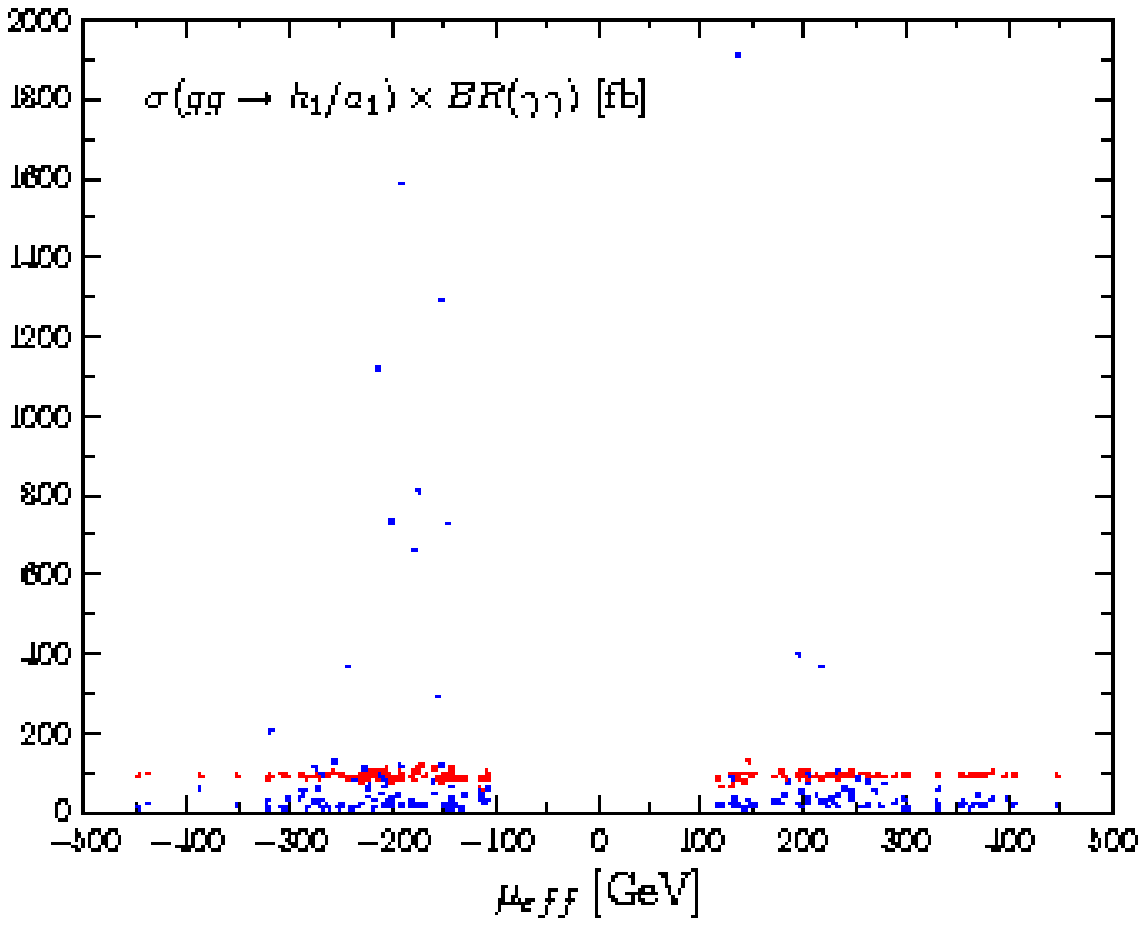}&\hspace*{-1.5truecm}\includegraphics[scale=0.35]{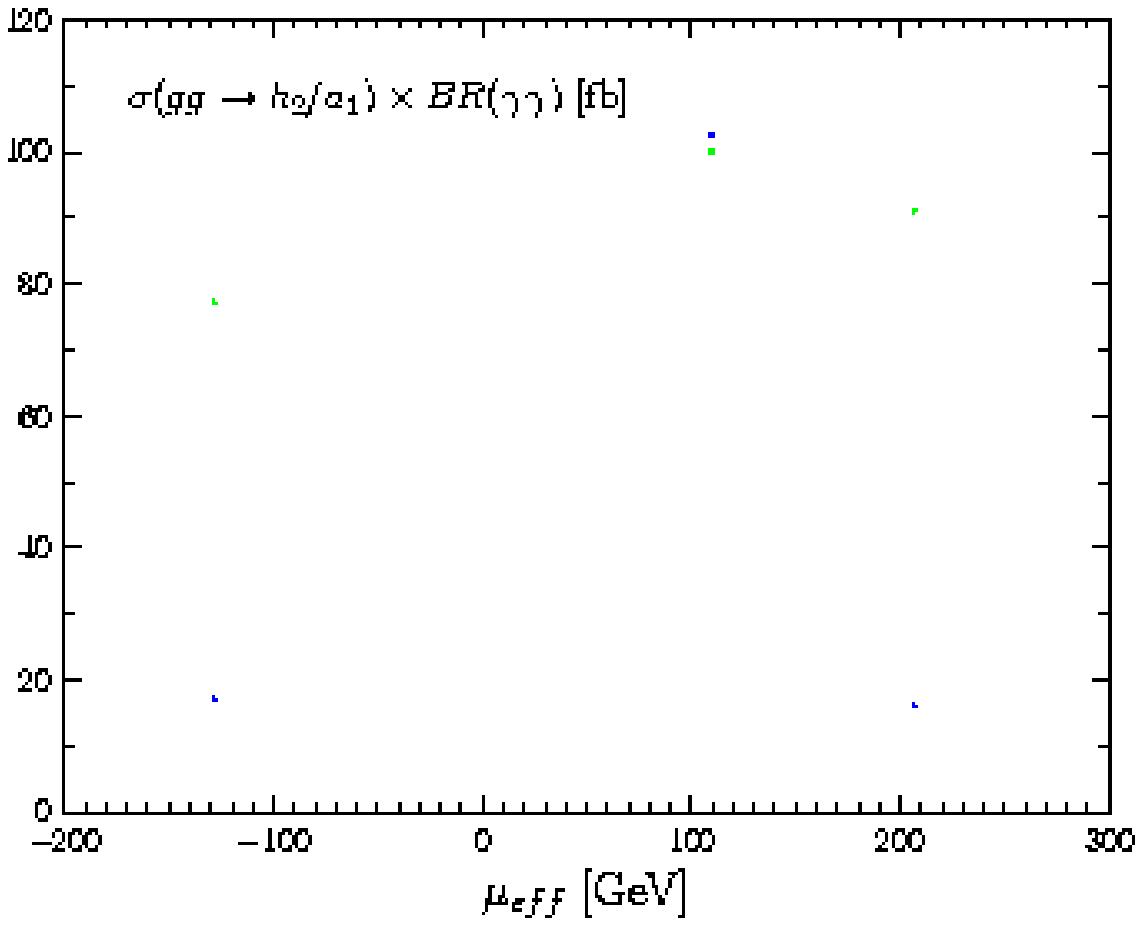}
\end{tabular}
\caption{Cross-section times BR of $h_1$ (red), $h_2$ (green) and $a_1$ (blue), when potentially visible individually and when two of these are potentially visible simultaneously, plotted against the parameter $\mu$.}
\end{figure}

\begin{figure}
\begin{tabular}{ccc}
\hspace*{-1.5truecm}\includegraphics[scale=0.35]{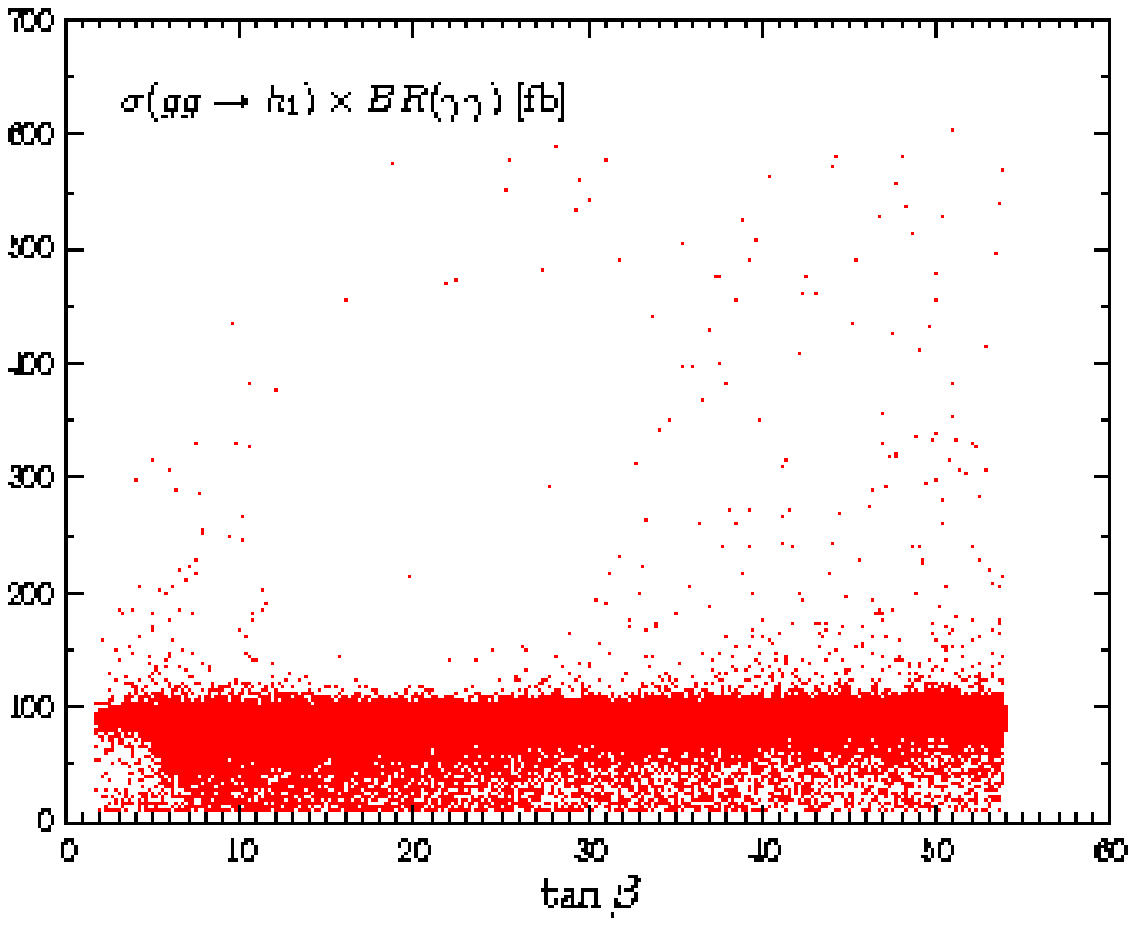}&\hspace*{-1.5truecm}\includegraphics[scale=0.35]{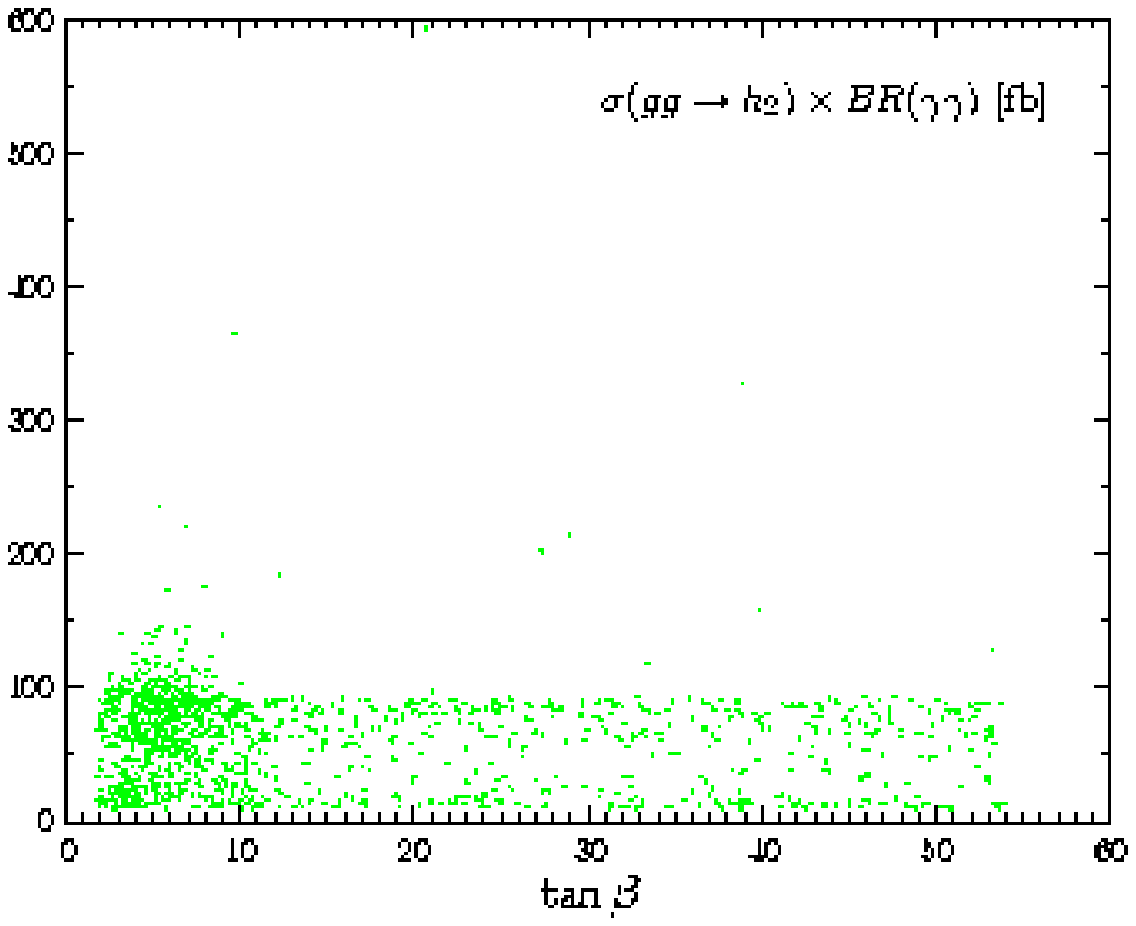} &\hspace*{-1.5truecm}\includegraphics[scale=0.35]{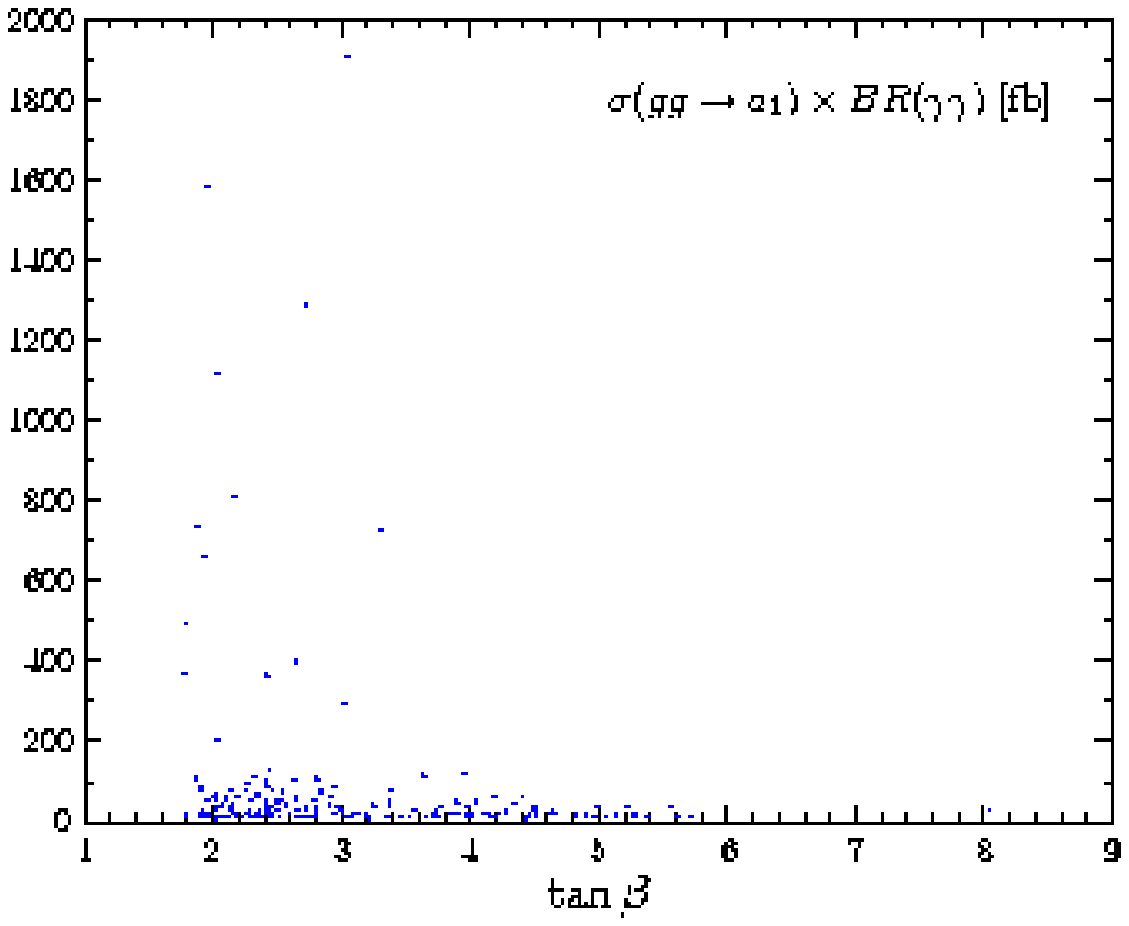}\\
\hspace*{-1.5truecm}\includegraphics[scale=0.35]{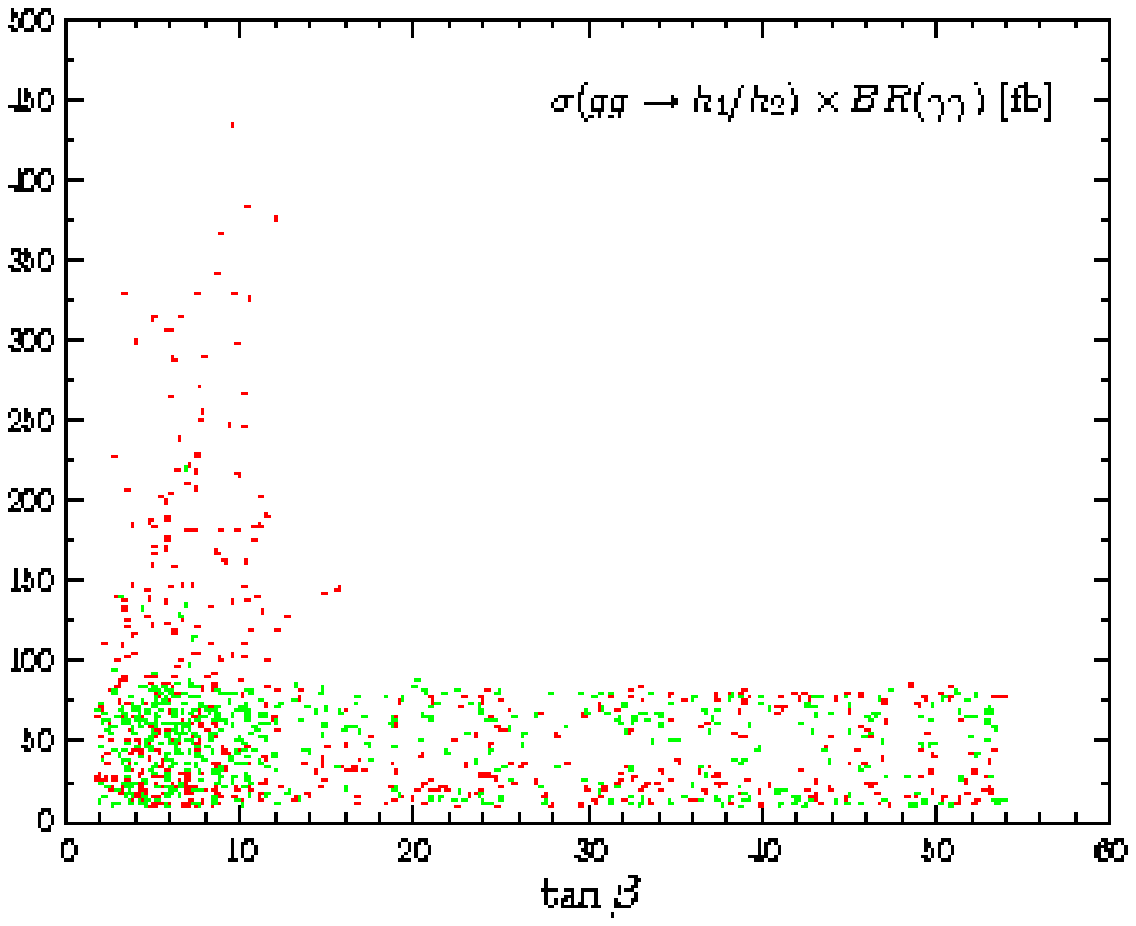}&\hspace*{-1.5truecm}\includegraphics[scale=0.35]{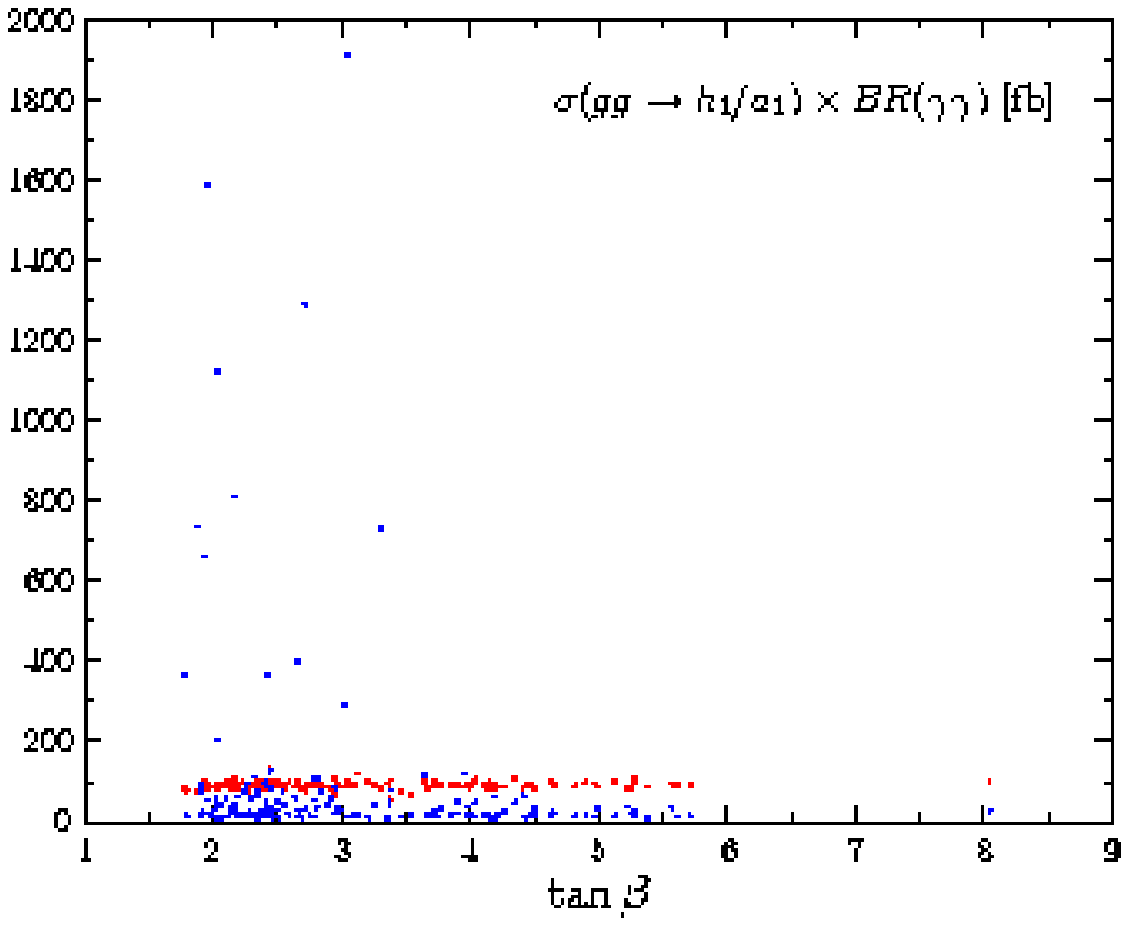}&\hspace*{-1.5truecm}\includegraphics[scale=0.35]{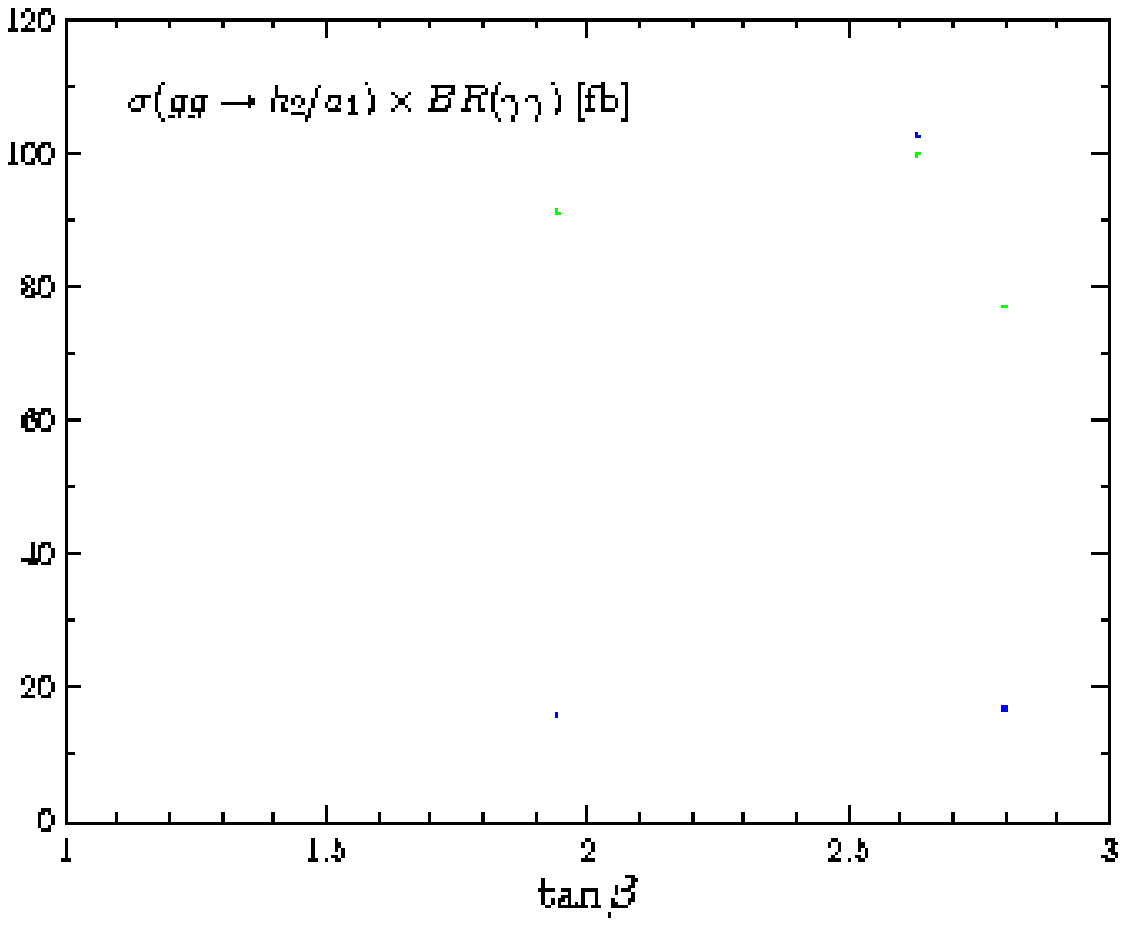}
\end{tabular}
\caption{Cross-section times BR of $h_1$ (red), $h_2$ (green) and $a_1$ (blue), when potentially visible individually and when two of these are potentially visible simultaneously, plotted against the parameter $\tan\beta$.}
\end{figure}

\clearpage

\begin{figure}
\begin{tabular}{ccc}
\hspace*{-1.5truecm}\includegraphics[scale=0.35]{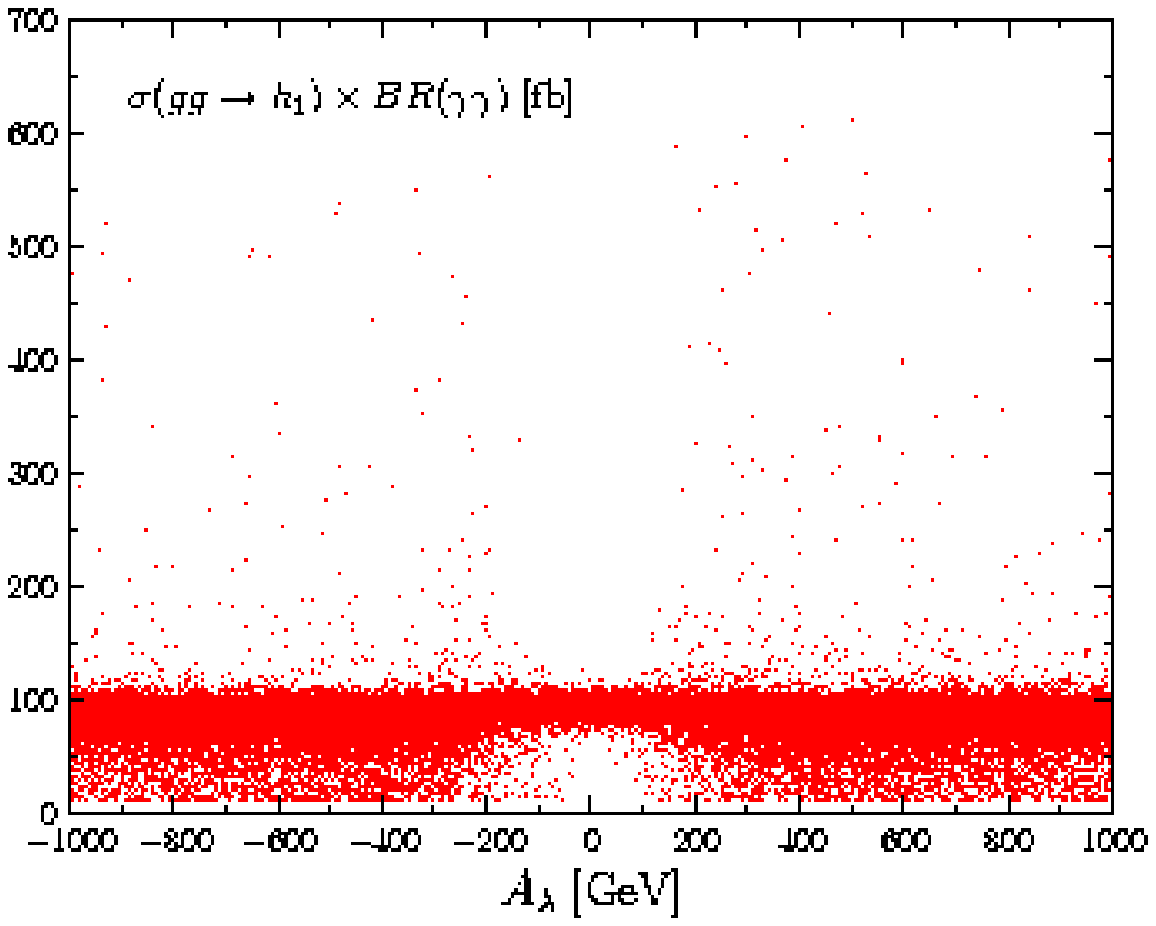}&\hspace*{-1.5truecm}\includegraphics[scale=0.35]{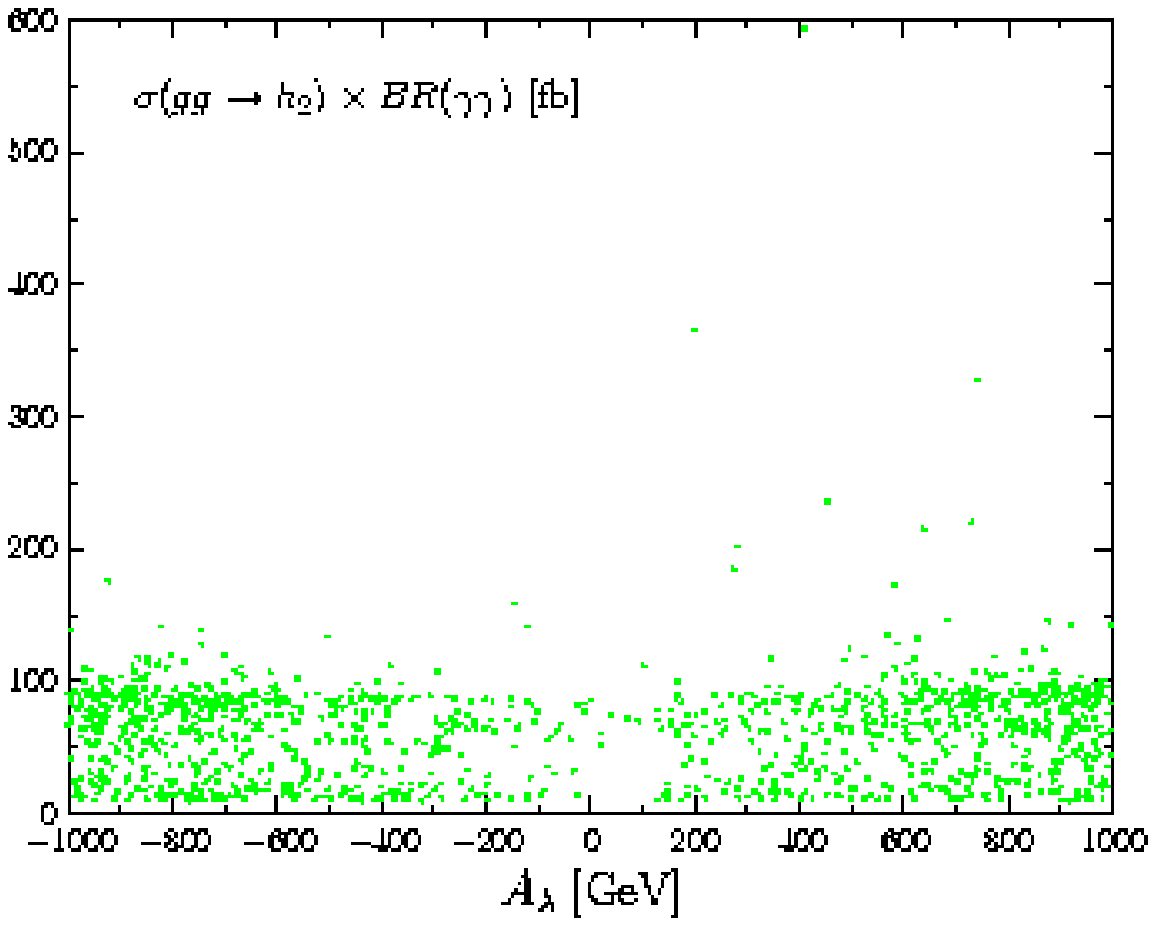} &\hspace*{-1.5truecm}\includegraphics[scale=0.35]{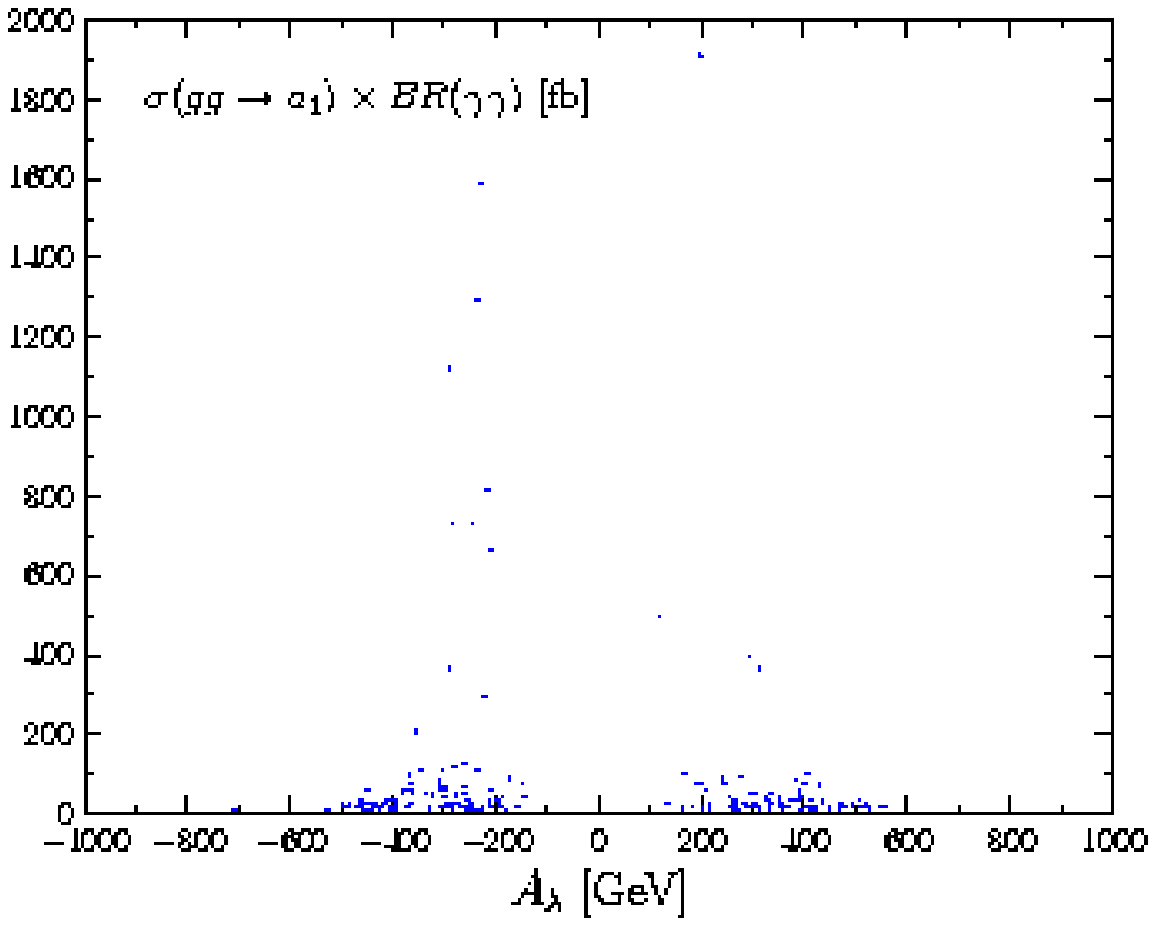}\\
\hspace*{-1.5truecm}\includegraphics[scale=0.35]{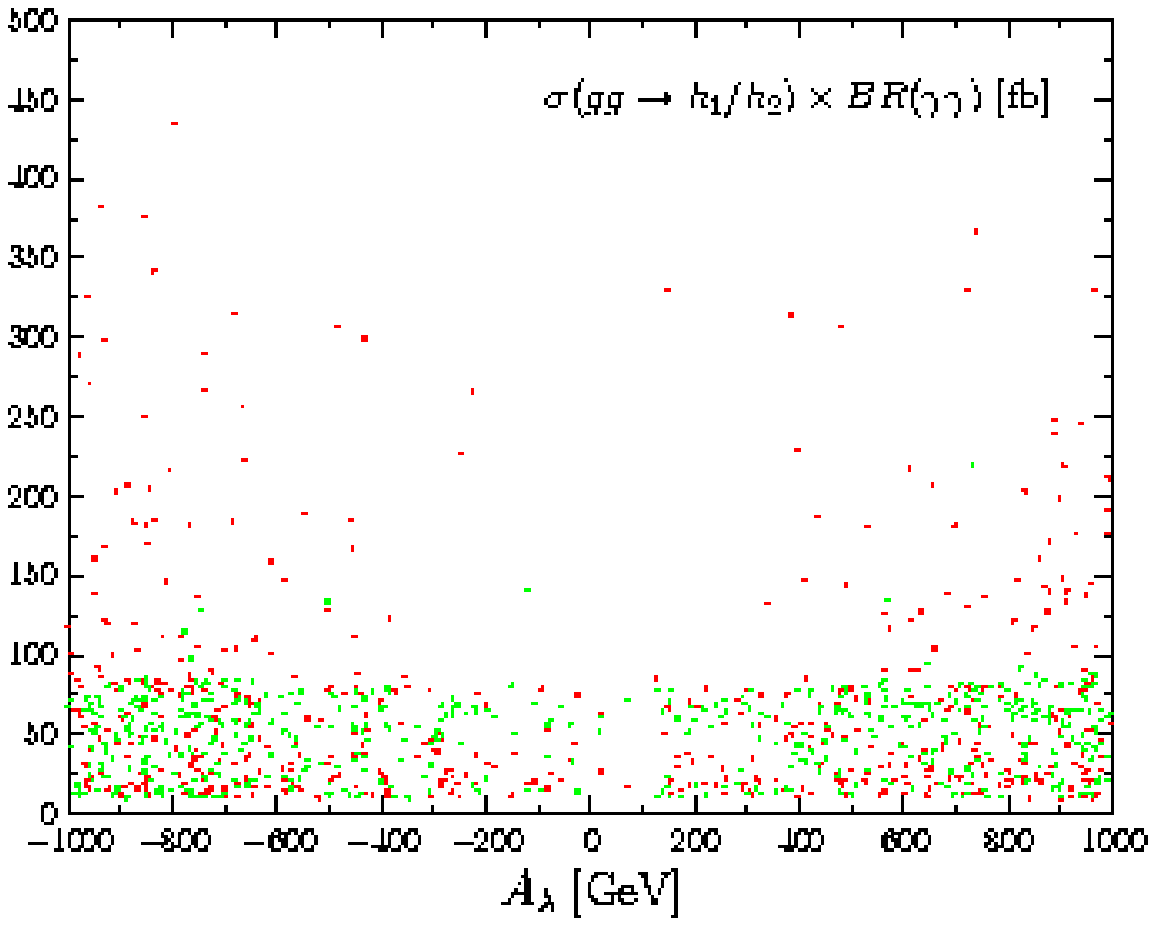}&\hspace*{-1.5truecm}\includegraphics[scale=0.35]{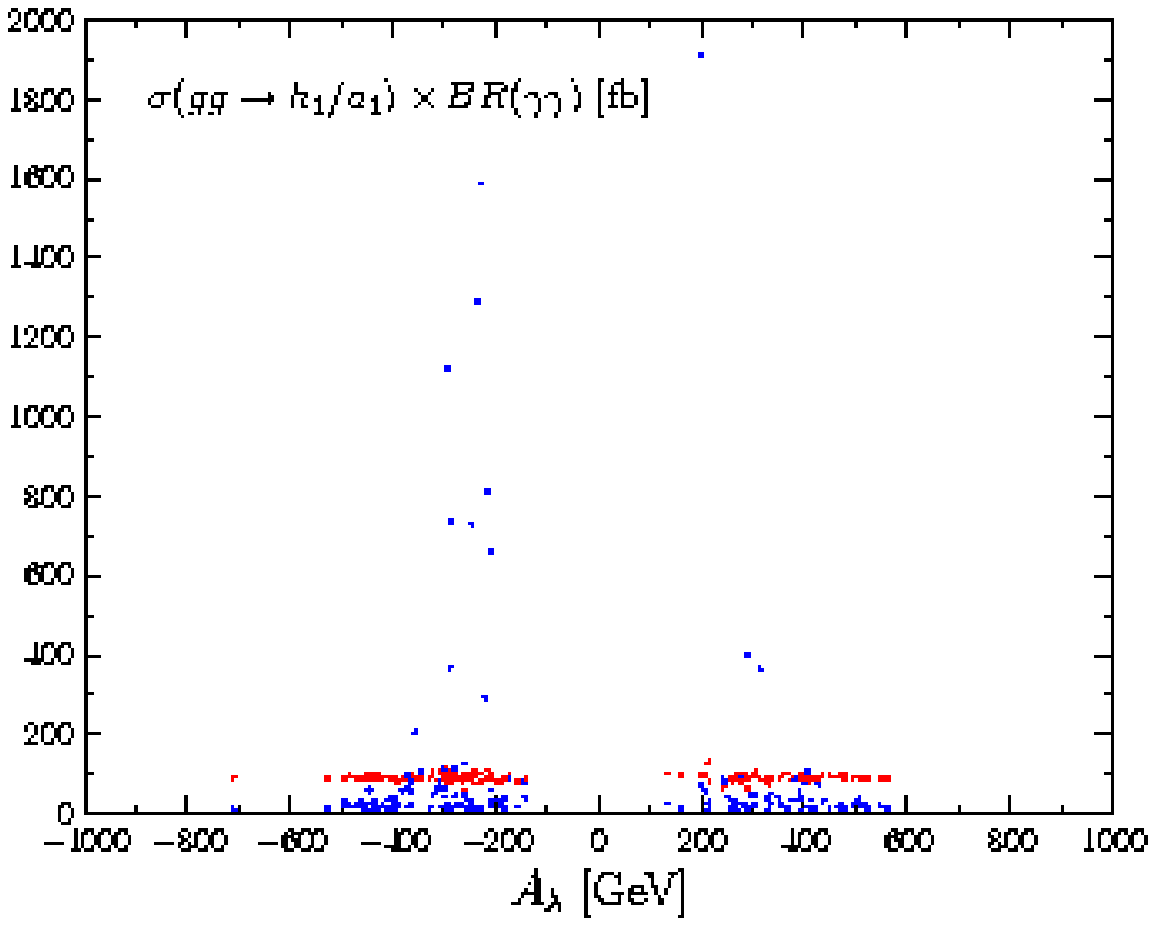}&\hspace*{-1.5truecm}\includegraphics[scale=0.35]{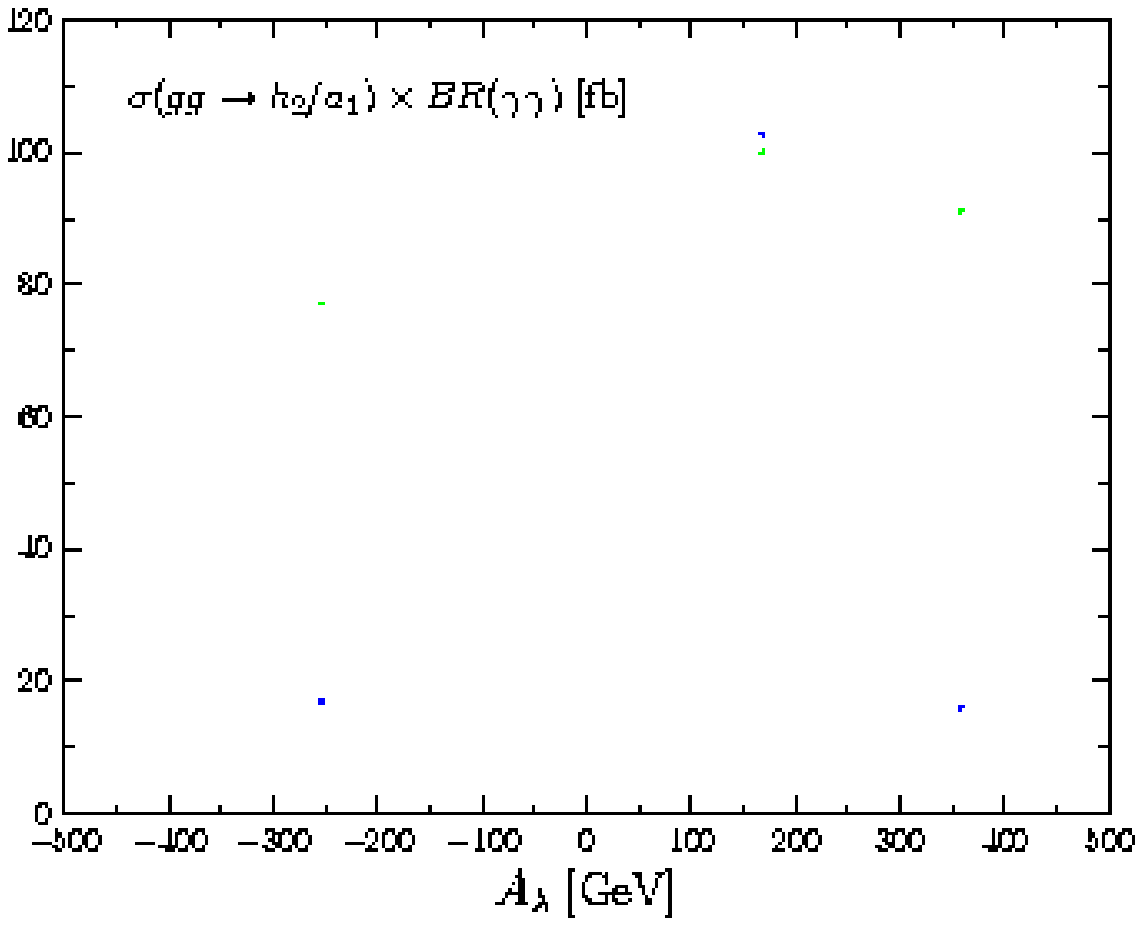}
\end{tabular}
\caption{Cross-section times BR of $h_1$ (red), $h_2$ (green) and $a_1$ (blue), when potentially visible individually and when two of these are potentially visible simultaneously, plotted against the parameter $A_{\lambda}$.}
\end{figure}

\begin{figure}
\begin{tabular}{ccc}
\hspace*{-1.5truecm}\includegraphics[scale=0.35]{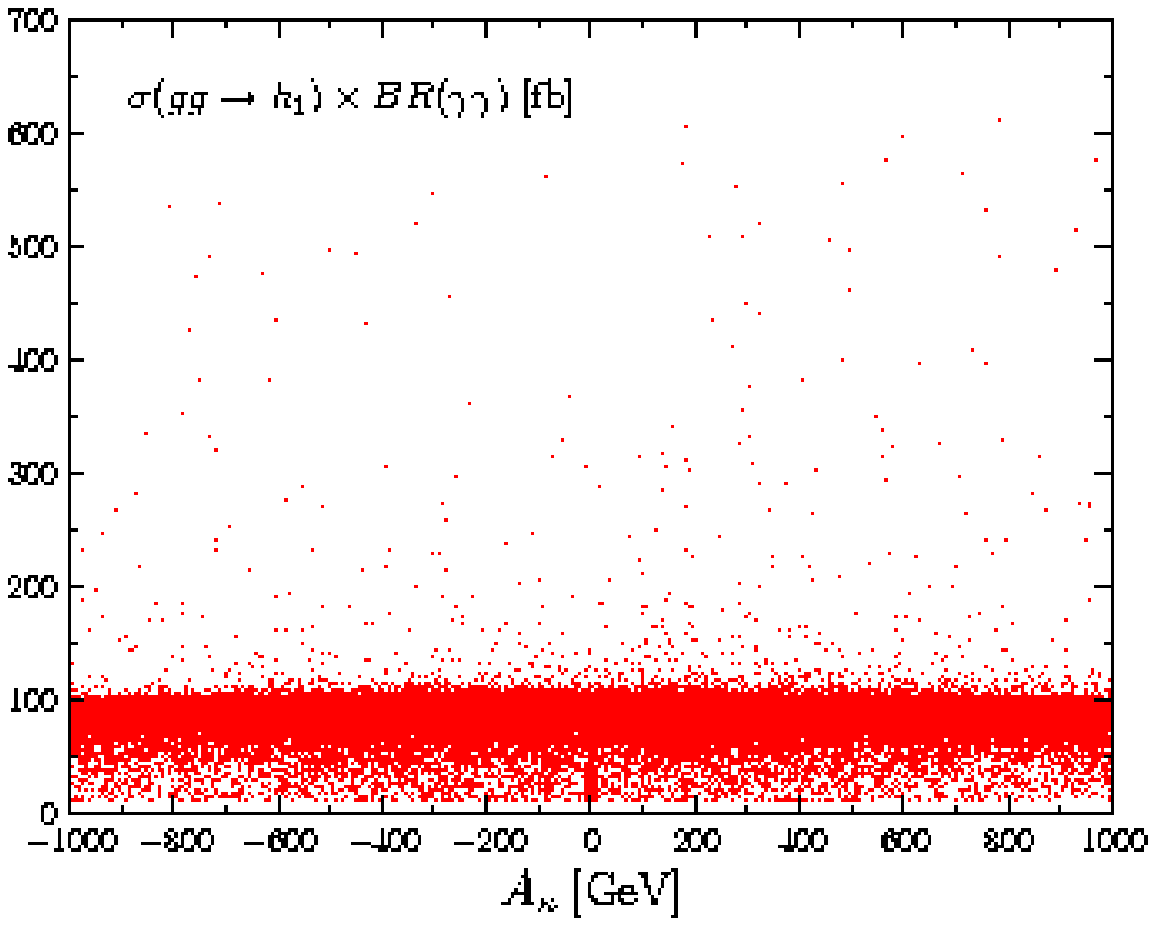}&\hspace*{-1.5truecm}\includegraphics[scale=0.35]{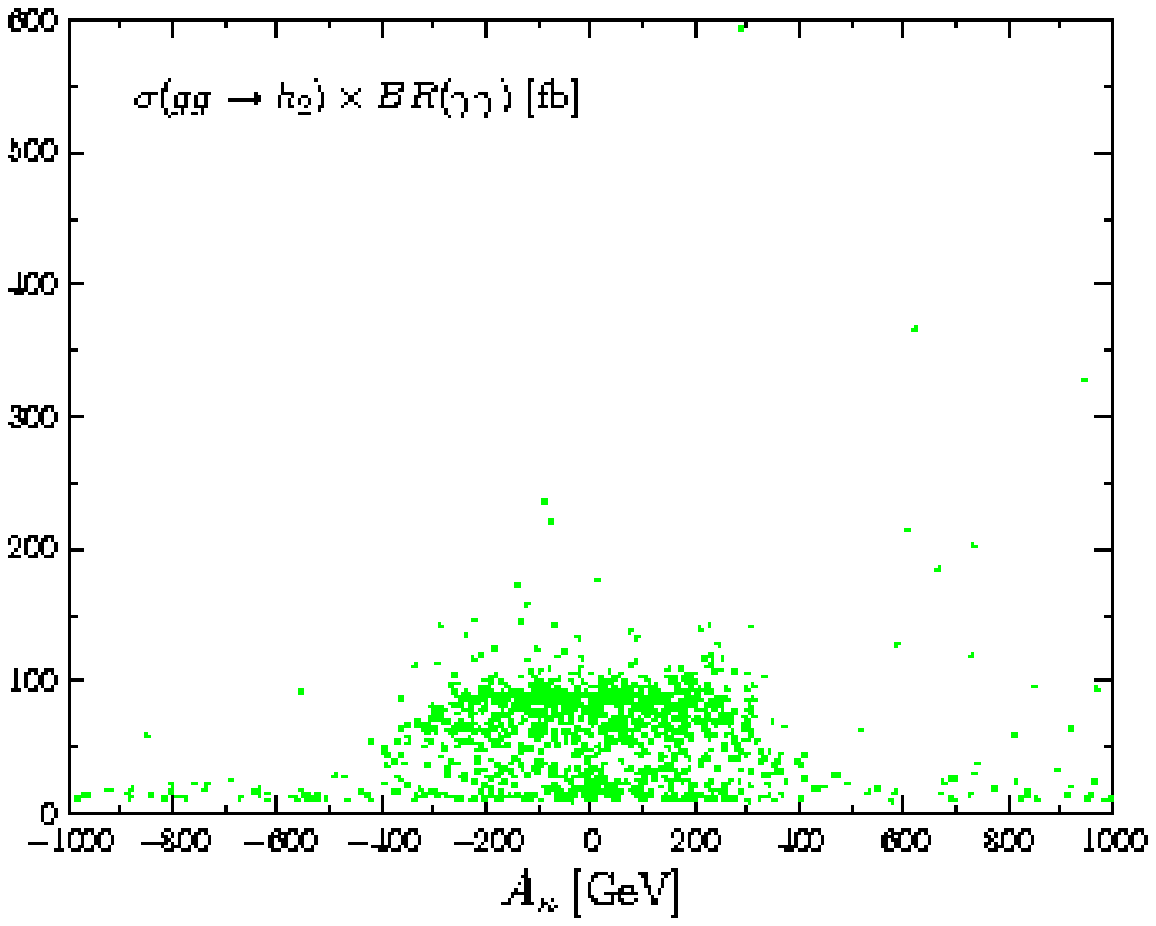} &\hspace*{-1.5truecm}\includegraphics[scale=0.35]{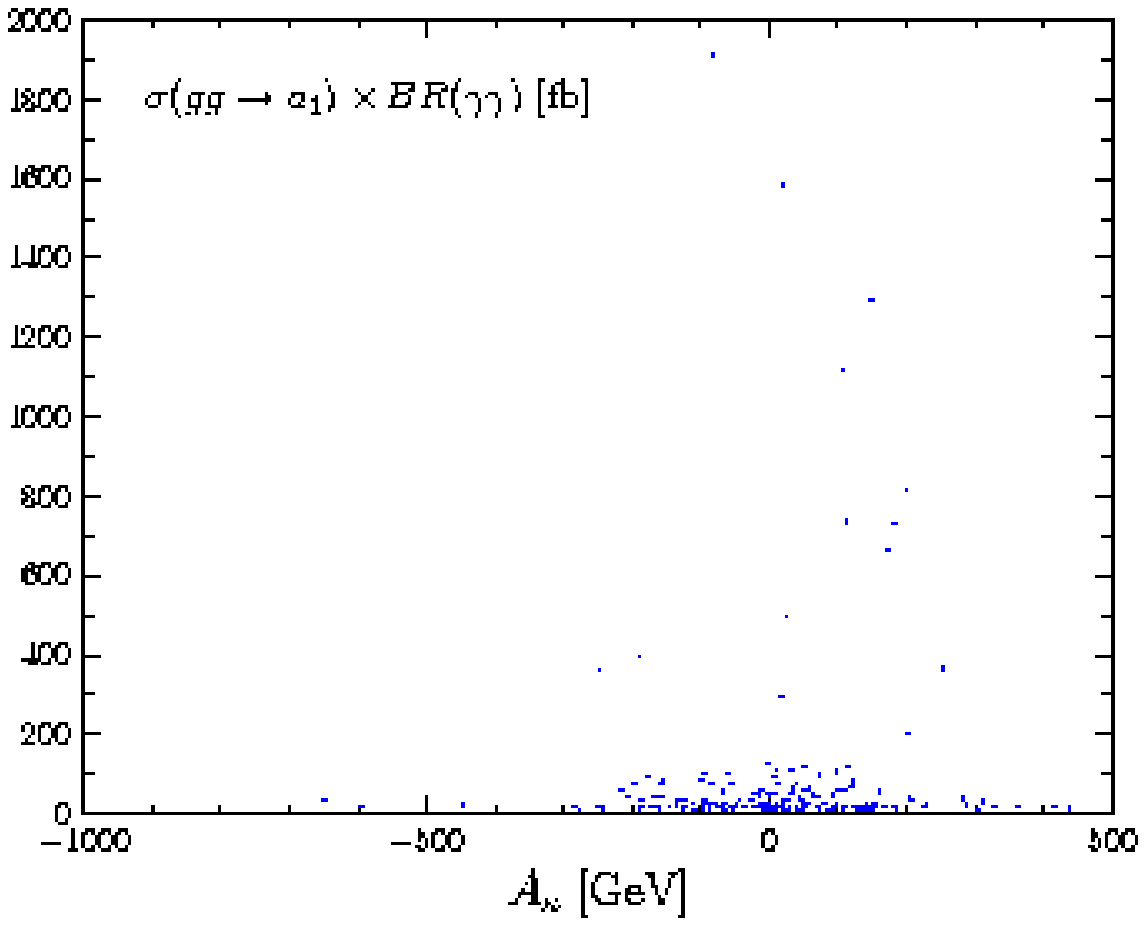}\\
\hspace*{-1.5truecm}\includegraphics[scale=0.35]{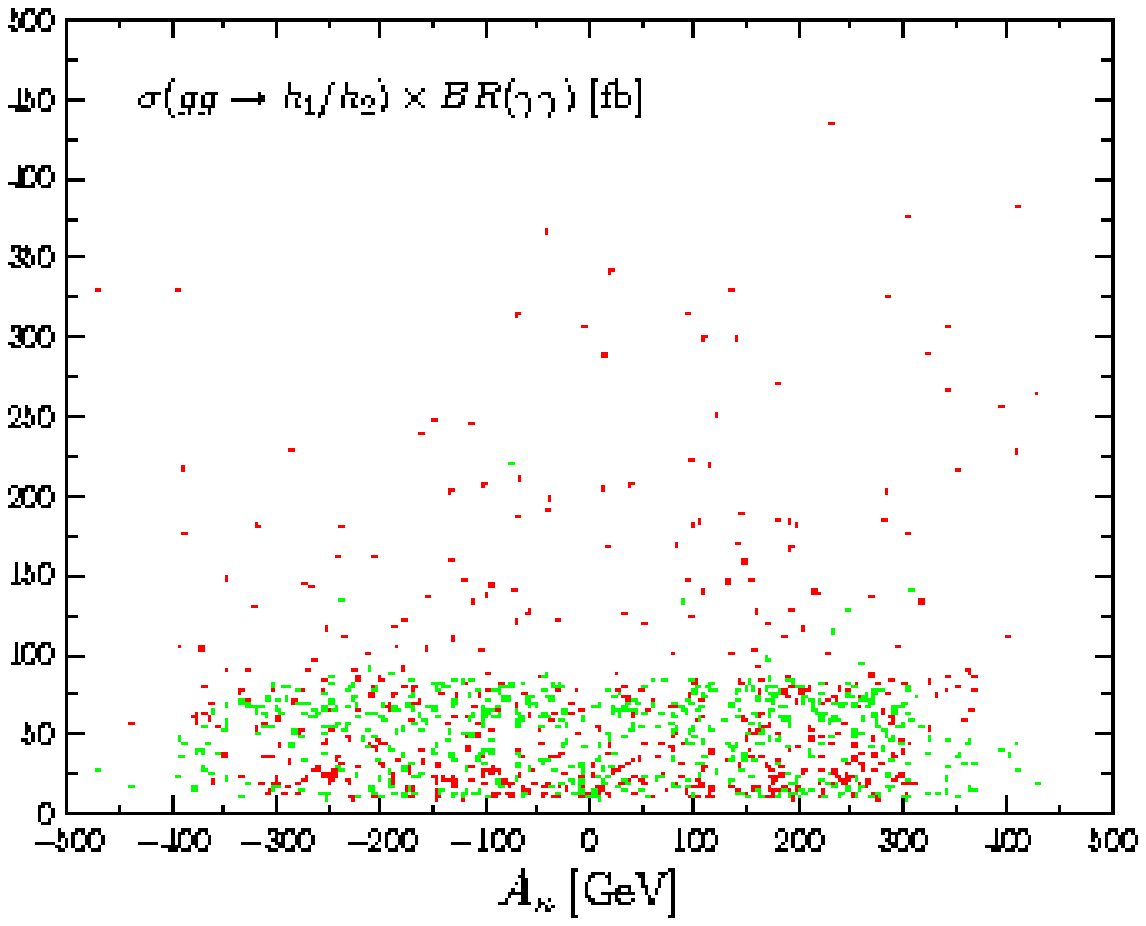}&\hspace*{-1.5truecm}\includegraphics[scale=0.35]{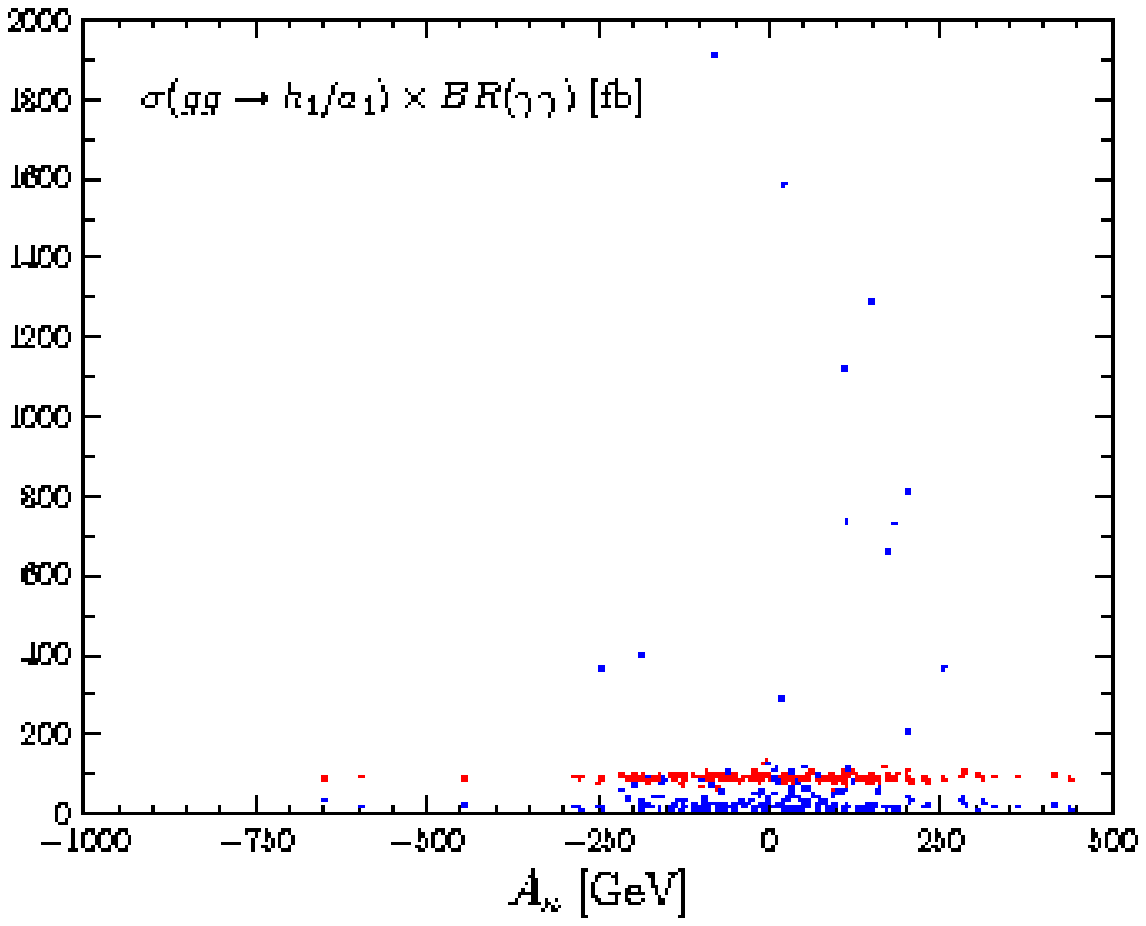}&\hspace*{-1.5truecm}\includegraphics[scale=0.35]{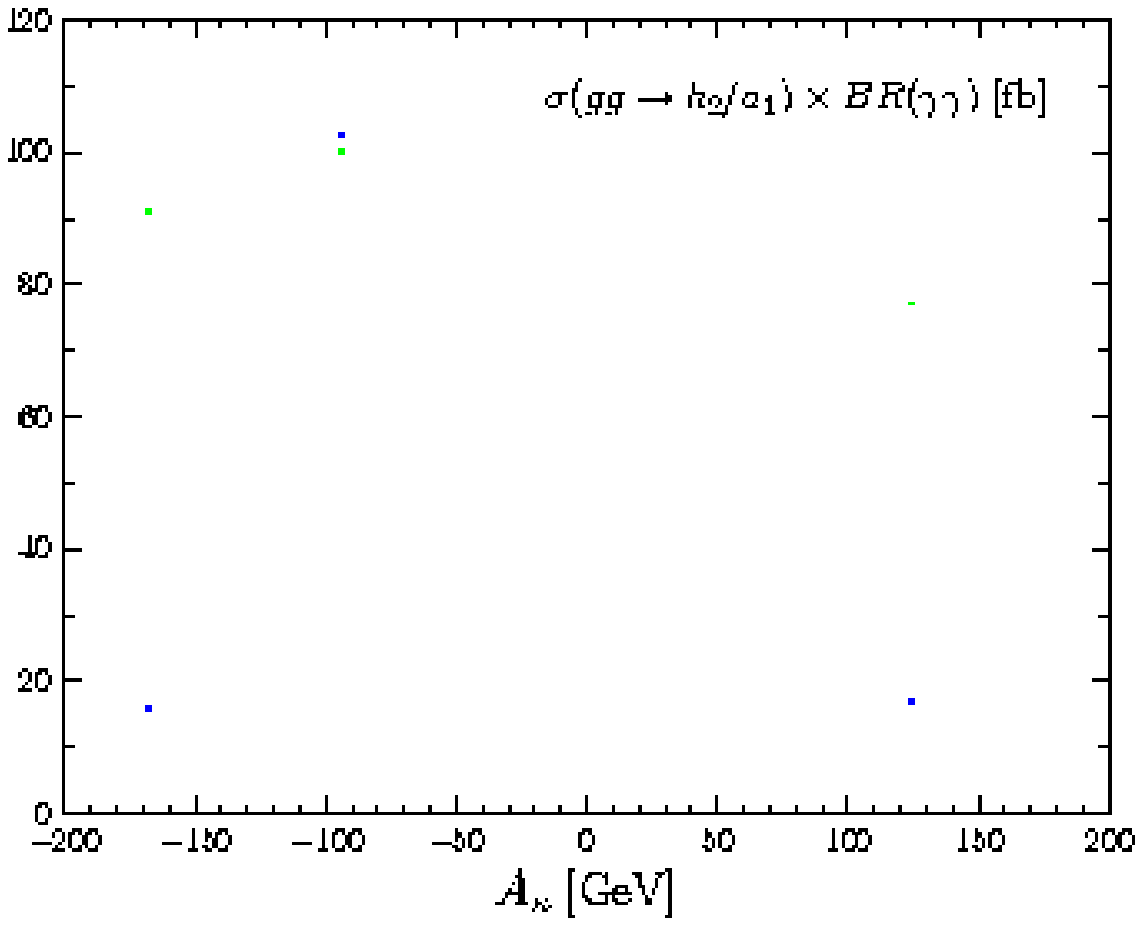}
\end{tabular}
\caption{Cross-section times BR of $h_1$ (red), $h_2$ (green) and $a_1$ (blue), when potentially visible individually and when two of these are potentially visible simultaneously, plotted against the parameter $A_{\kappa}$.}
\end{figure}

\clearpage

\begin{figure}
\begin{tabular}{ccc}
\hspace*{-1.5truecm}\includegraphics[scale=0.35]{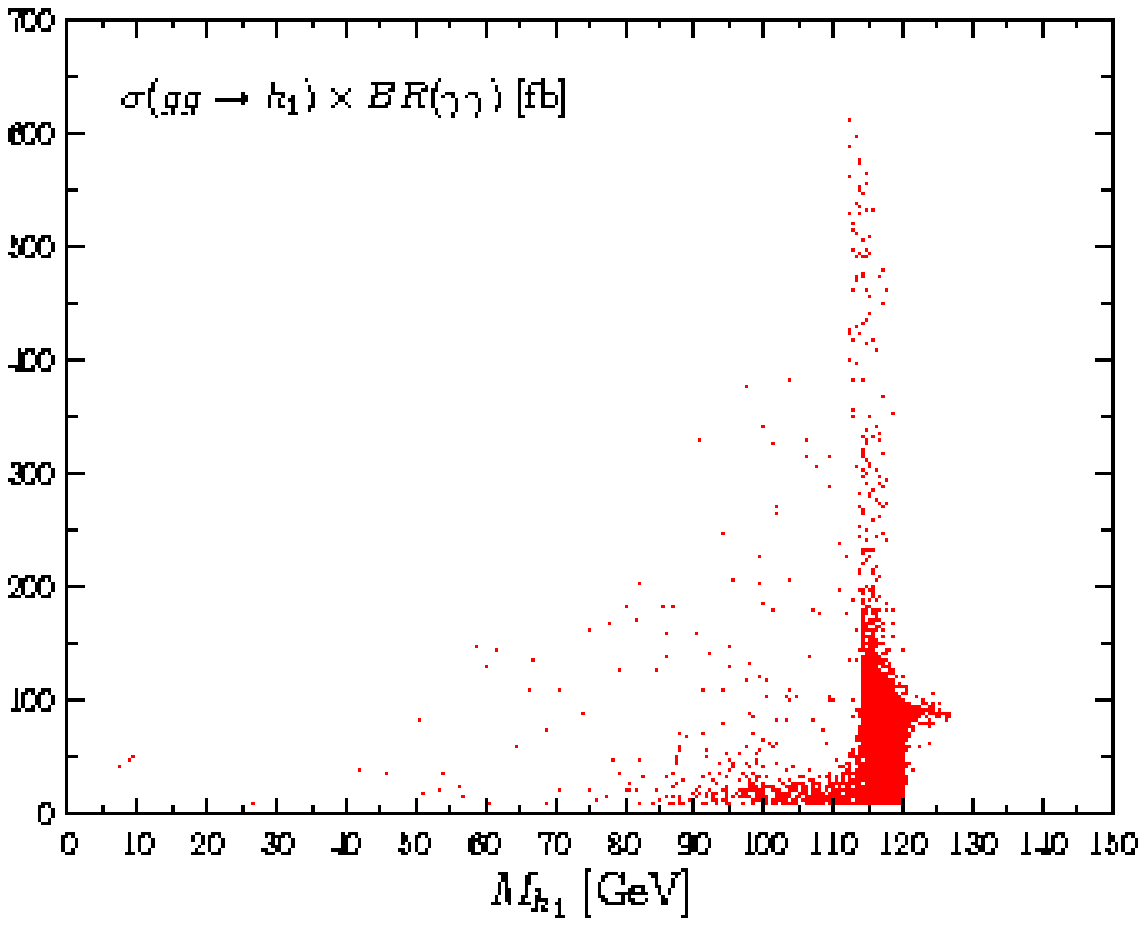}&\hspace*{-1.5truecm}\includegraphics[scale=0.35]{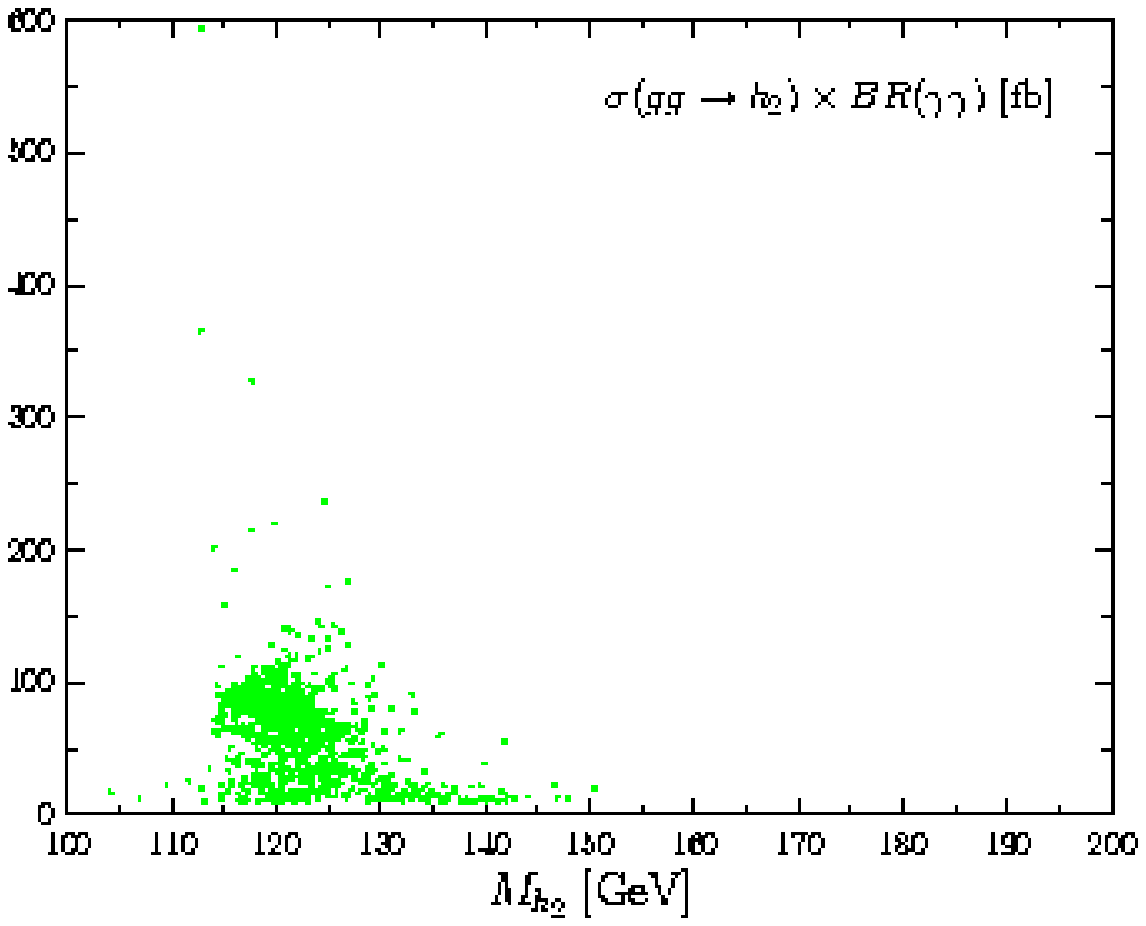} &\hspace*{-1.5truecm}\includegraphics[scale=0.35]{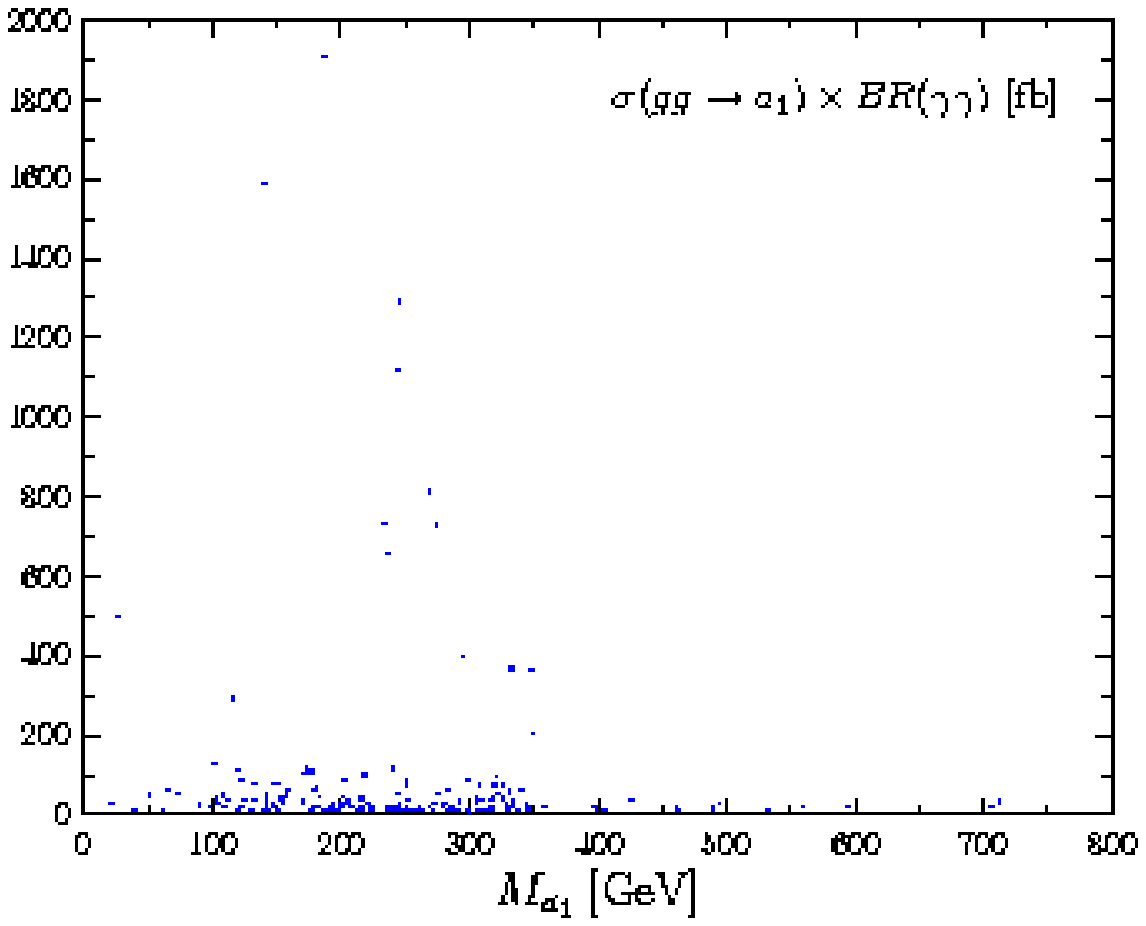}\\
\hspace*{-1.5truecm}\includegraphics[scale=0.35]{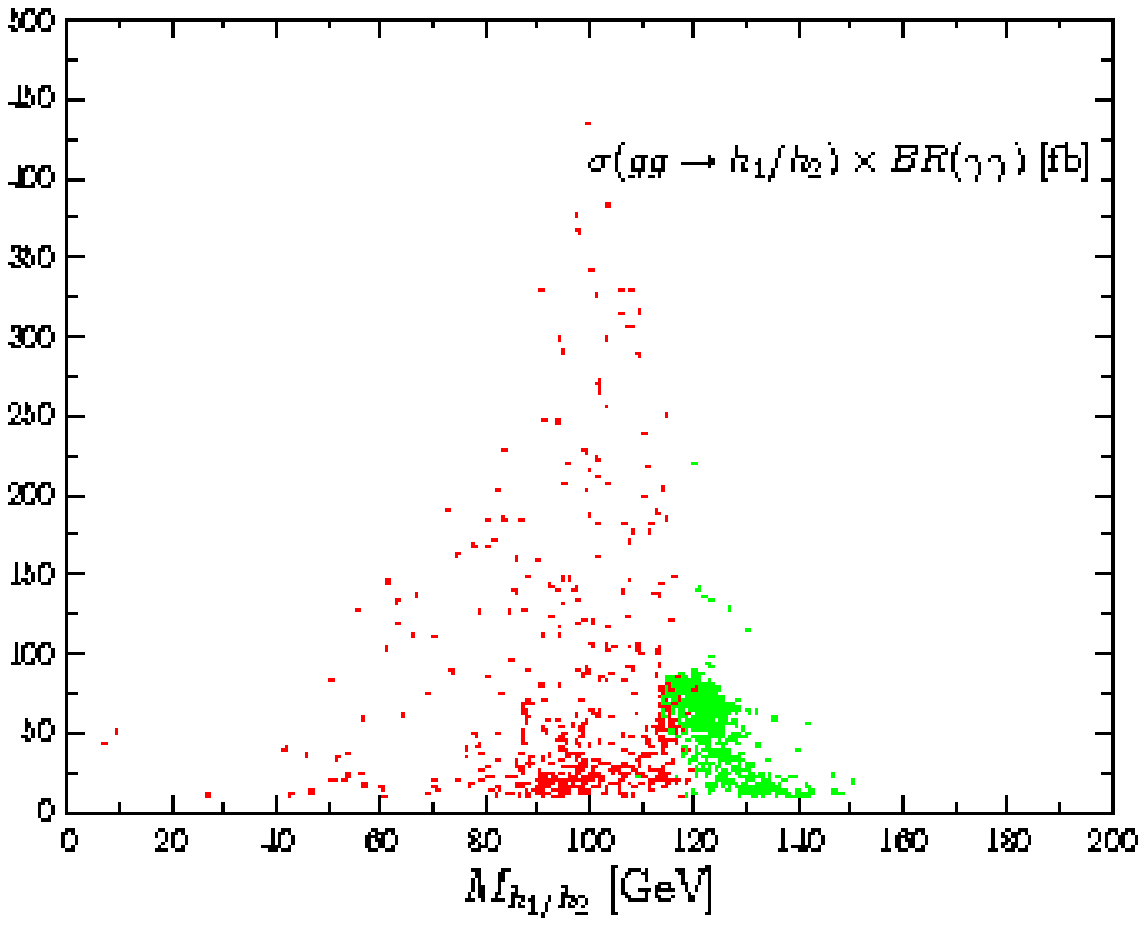}&\hspace*{-1.5truecm}\includegraphics[scale=0.35]{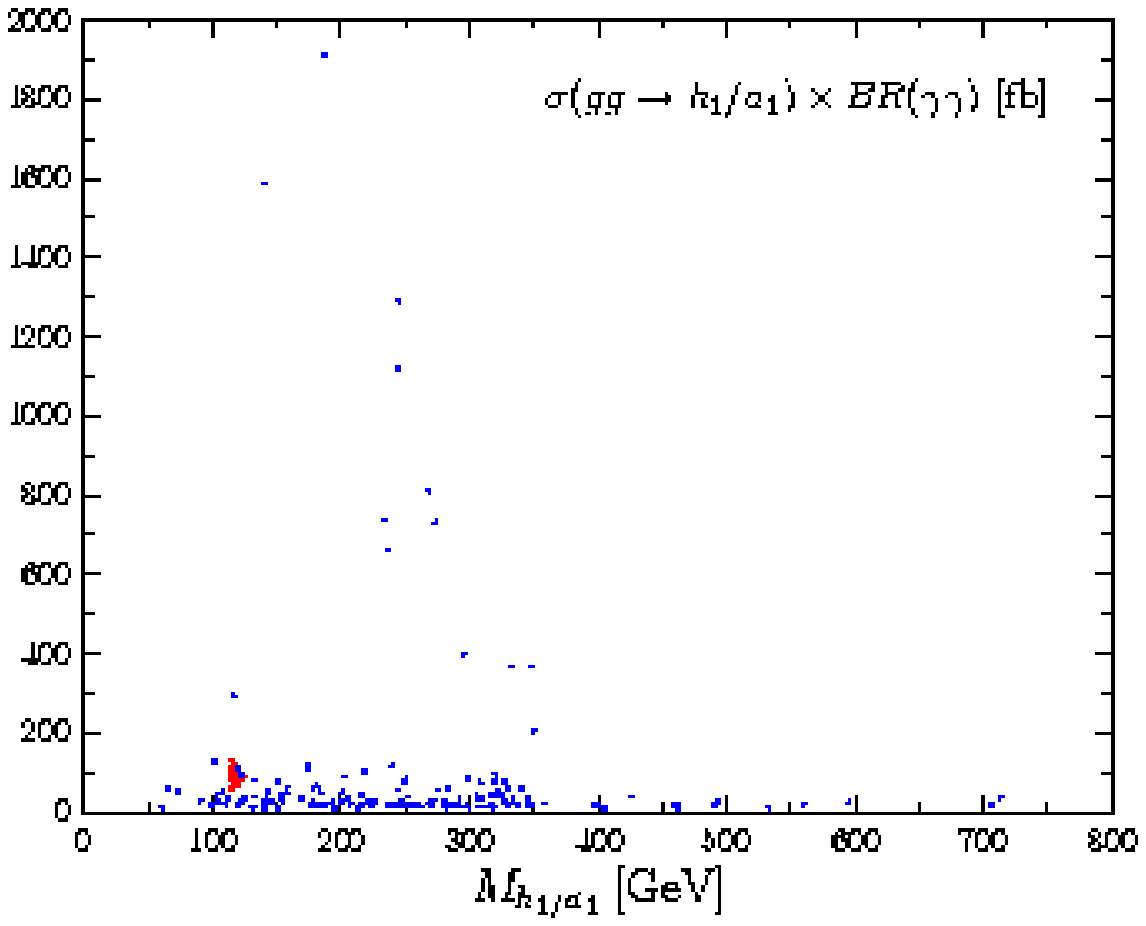}&\hspace*{-1.5truecm}\includegraphics[scale=0.35]{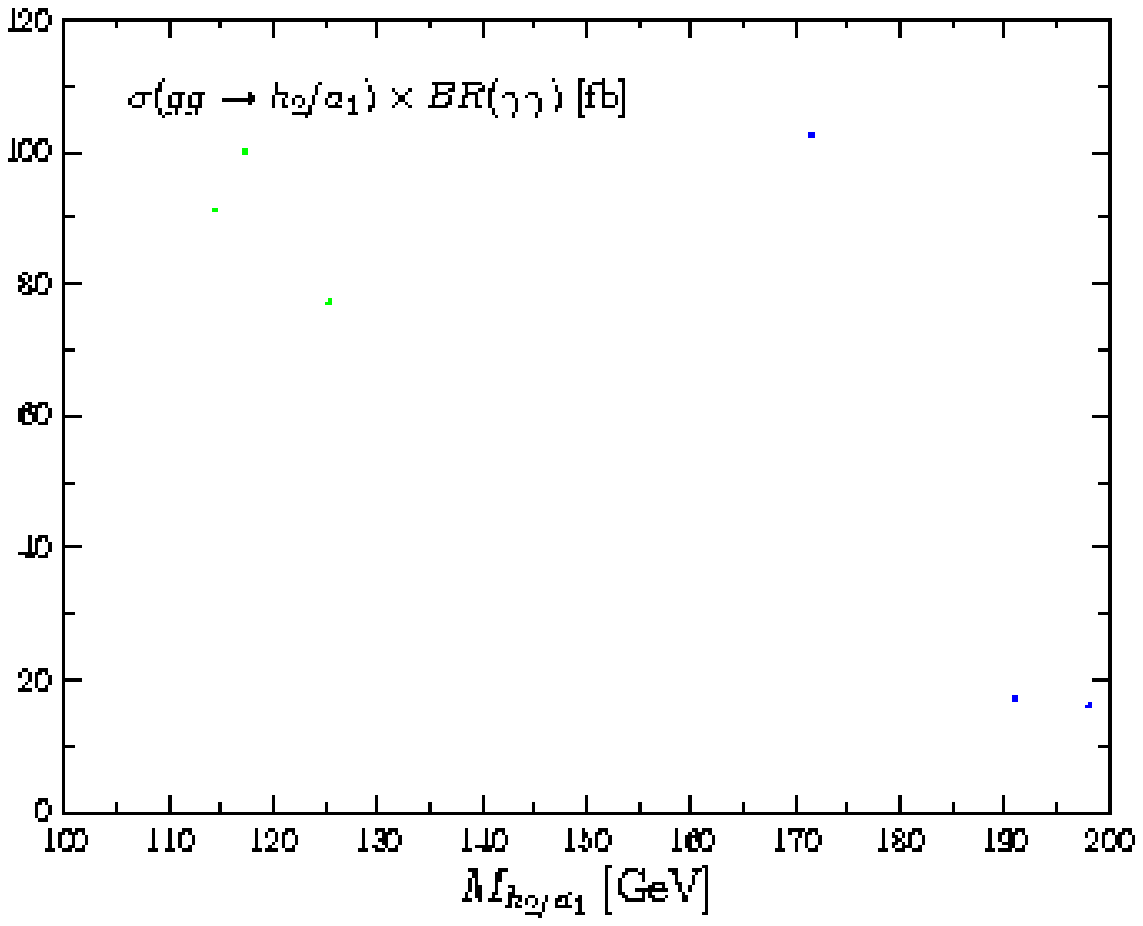}
\end{tabular}
\caption{Cross-section times BR of $h_1$ (red), $h_2$ (green) and $a_1$ (blue), when potentially visible individually and when two of these are potentially visible simultaneously, plotted against their respective masses.}
\end{figure}

\begin{figure}
\vspace*{-5.0truecm}
\centering\includegraphics[scale=0.9]{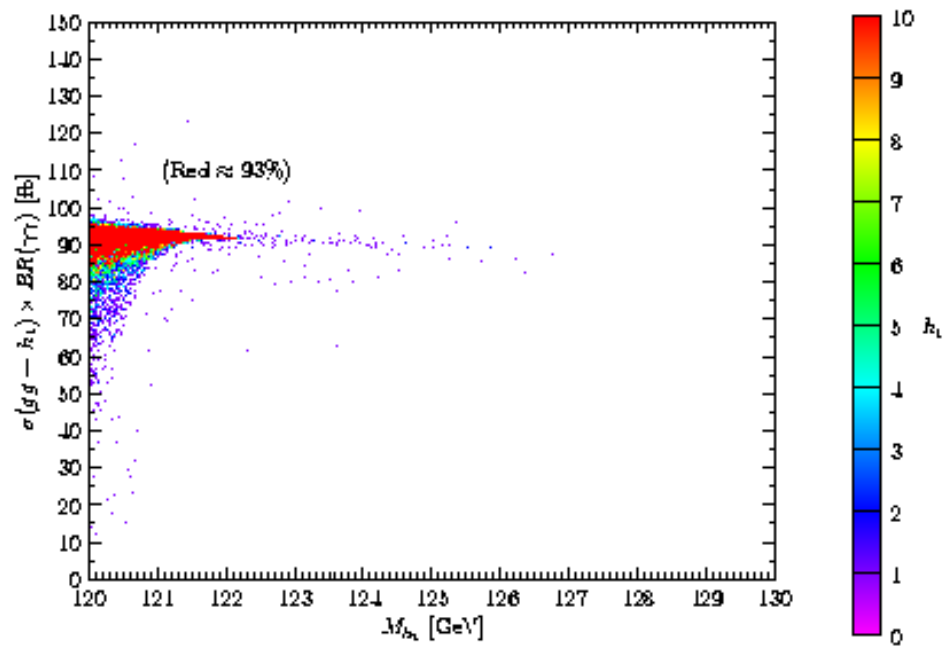}
\vspace*{-5.0truecm}
\caption{The distribution of points where one has only one potentially visible NMSSM $h_1$ state with mass beyond the 
MSSM upper mass limit on the corresponding Higgs state. The scale on the right represents a measure of
density of the points.}
\end{figure}

\clearpage

\begin{figure}
\centering\begin{tabular}{cc}
\hspace*{-0.65truecm}\includegraphics[scale=0.45]{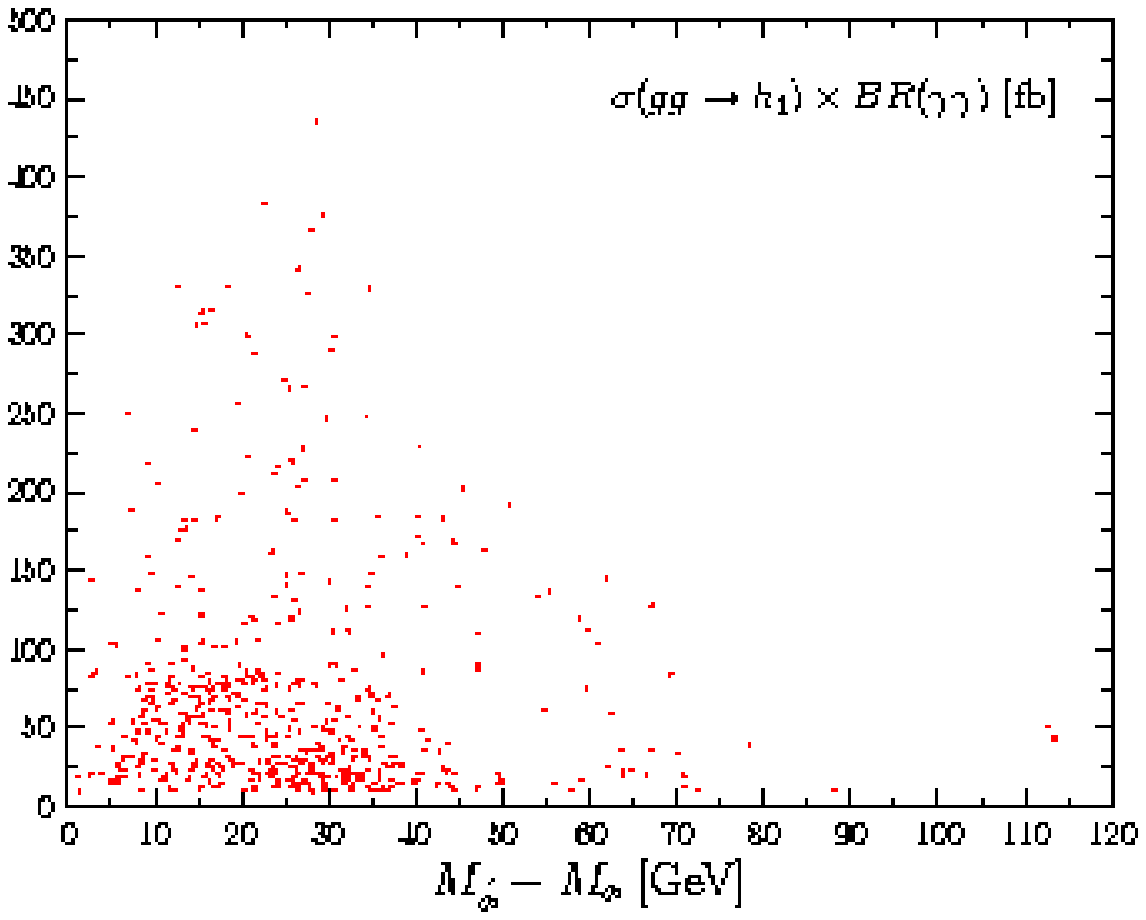}&\hspace*{-1.5truecm}\includegraphics[scale=0.45]{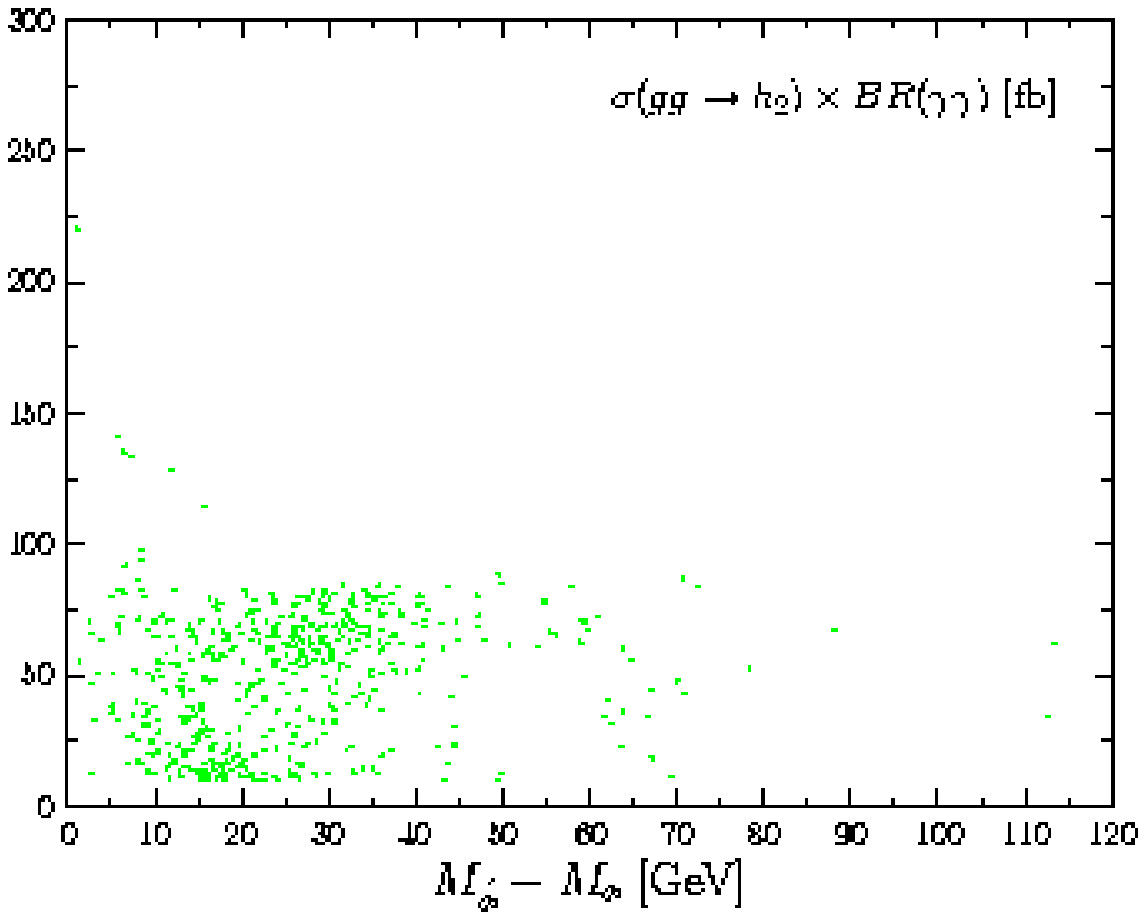}
\end{tabular}
\caption{Cross-section times BR of $h_1$ (red) and $h_2$ (green) plotted against their mass differences when the two are potentially visible simultaneously.}
\end{figure}

\begin{figure}
\centering\begin{tabular}{cc}
\hspace*{-0.65truecm}\includegraphics[scale=0.45]{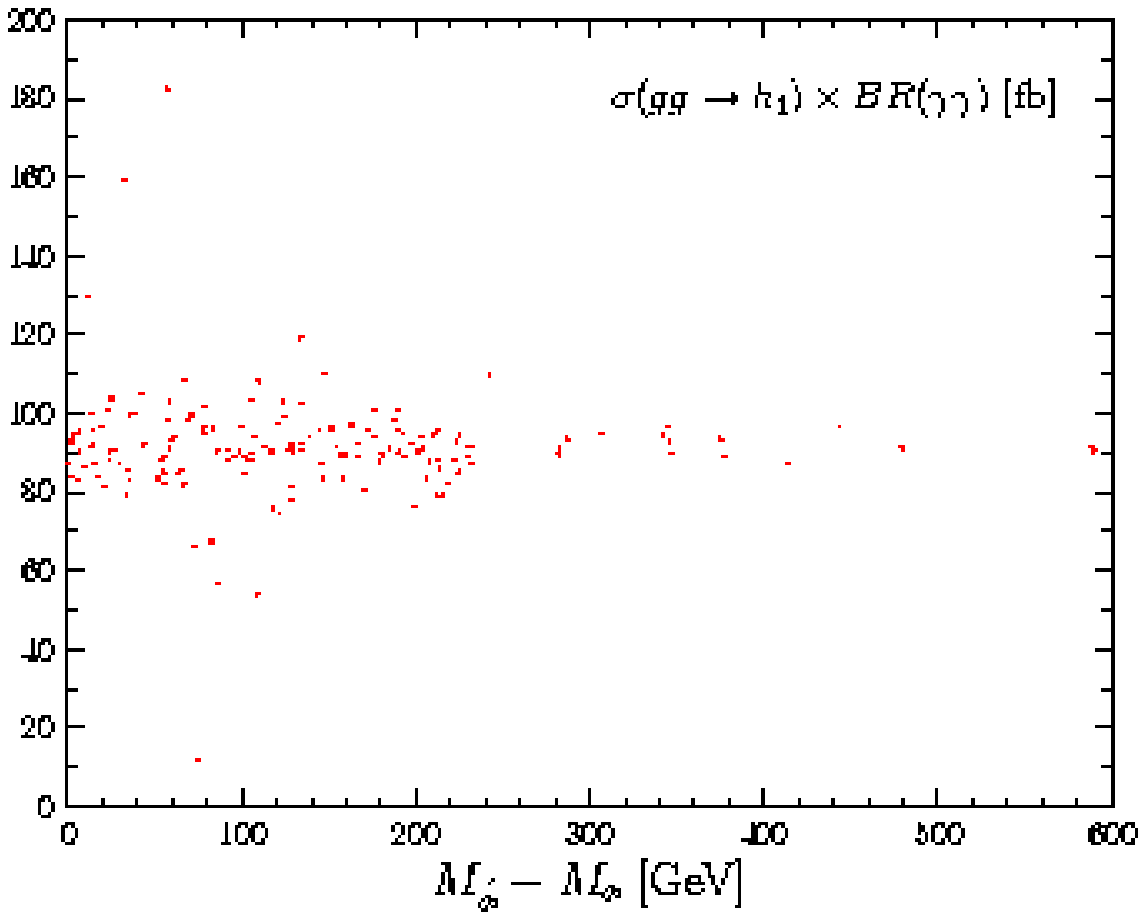}&\hspace*{-1.5truecm}\includegraphics[scale=0.45]{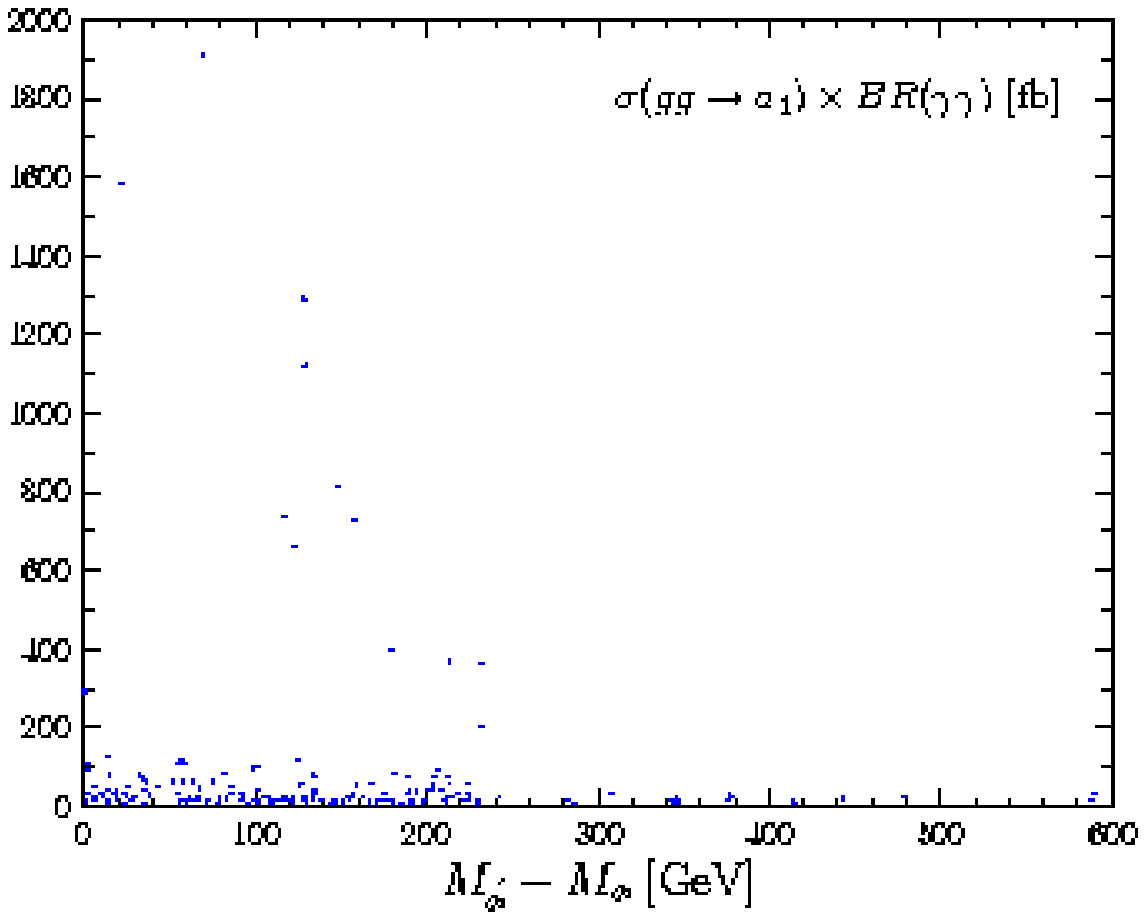}
\end{tabular}
\caption{Cross-section times BR of $h_1$ (red) and $a_1$ (blue) plotted against thier mass differences when the two are potentially visible simultaneously.}
\end{figure}

\begin{figure}
\centering\begin{tabular}{cc}
\hspace*{-0.65truecm}\includegraphics[scale=0.45]{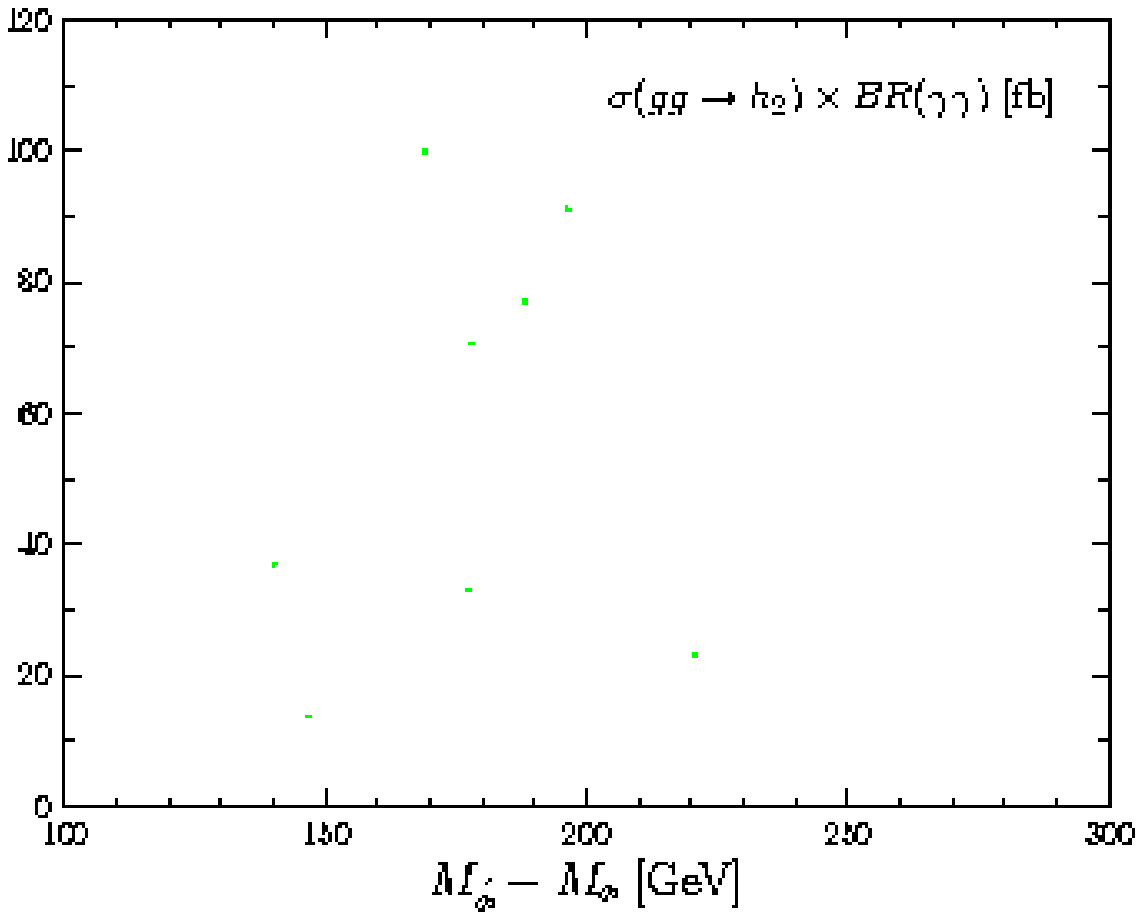}&\hspace*{-1.5truecm}\includegraphics[scale=0.45]{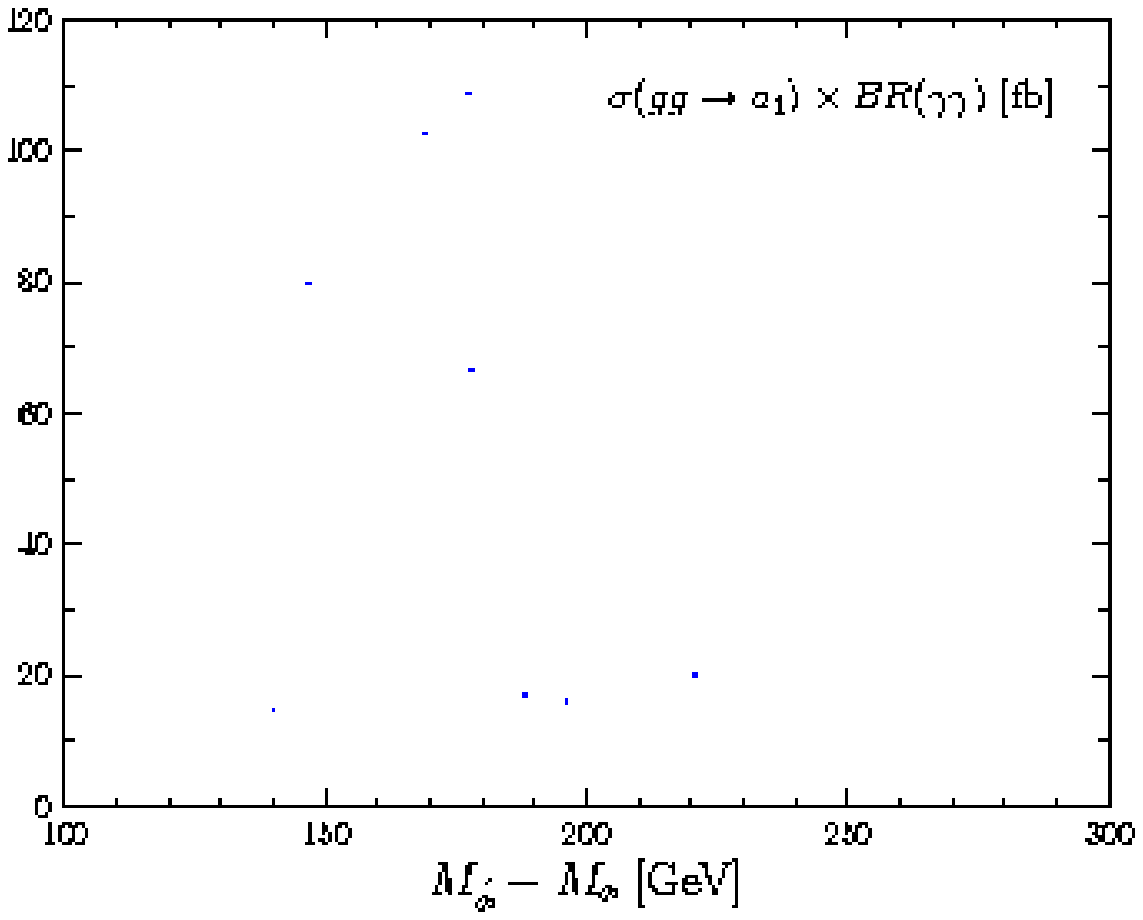}
\end{tabular}
\caption{Cross-section times BR of $h_2$ (green) and $a_1$ (blue) plotted against their mass differences when the two are potentially visible simultaneously.}
\end{figure}

\clearpage

\begin{figure}
\hspace*{3.5truecm}\includegraphics[scale=0.35,angle=90]{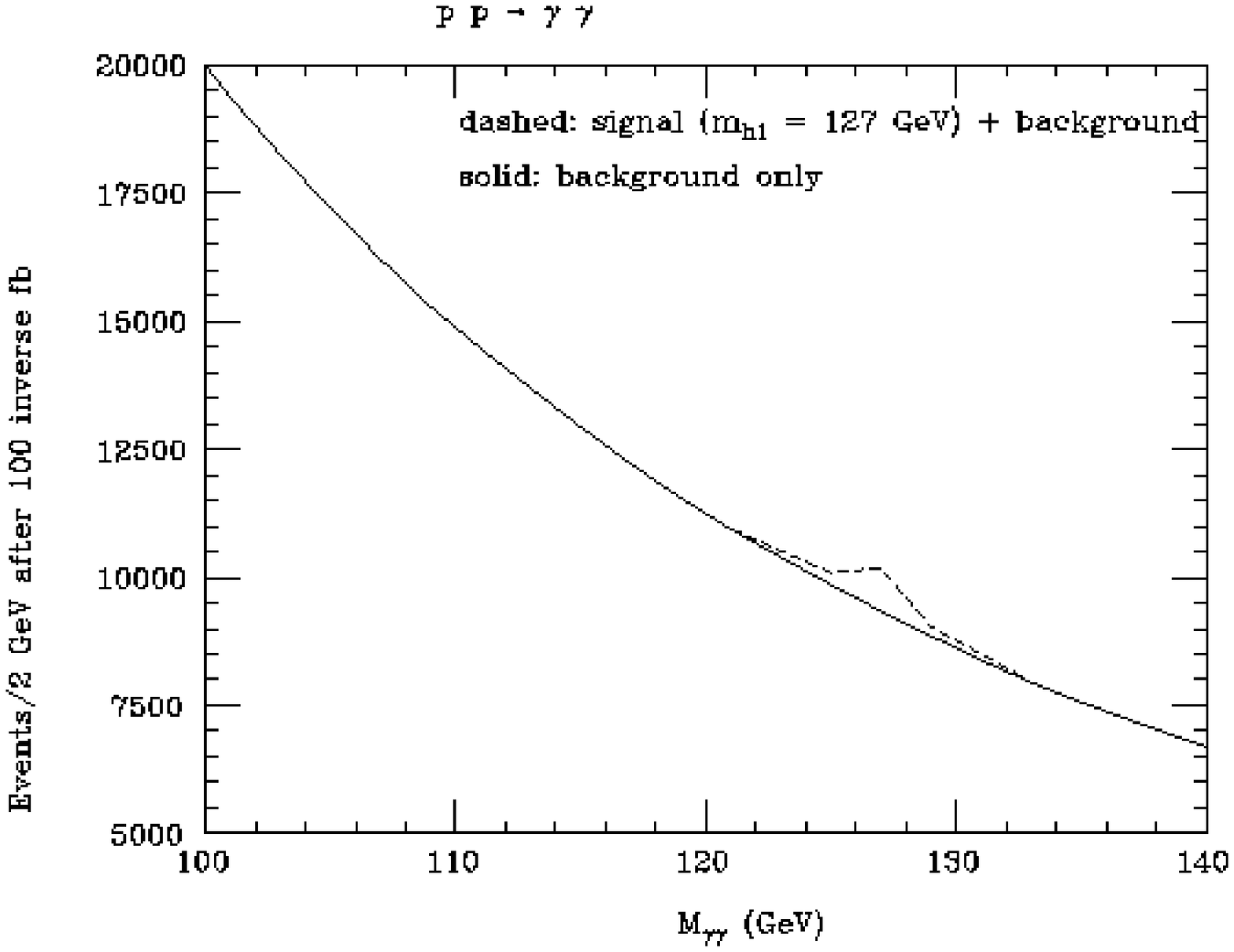}\\[-1.5cm]
\hspace*{3.5truecm}\includegraphics[scale=0.35,angle=90]{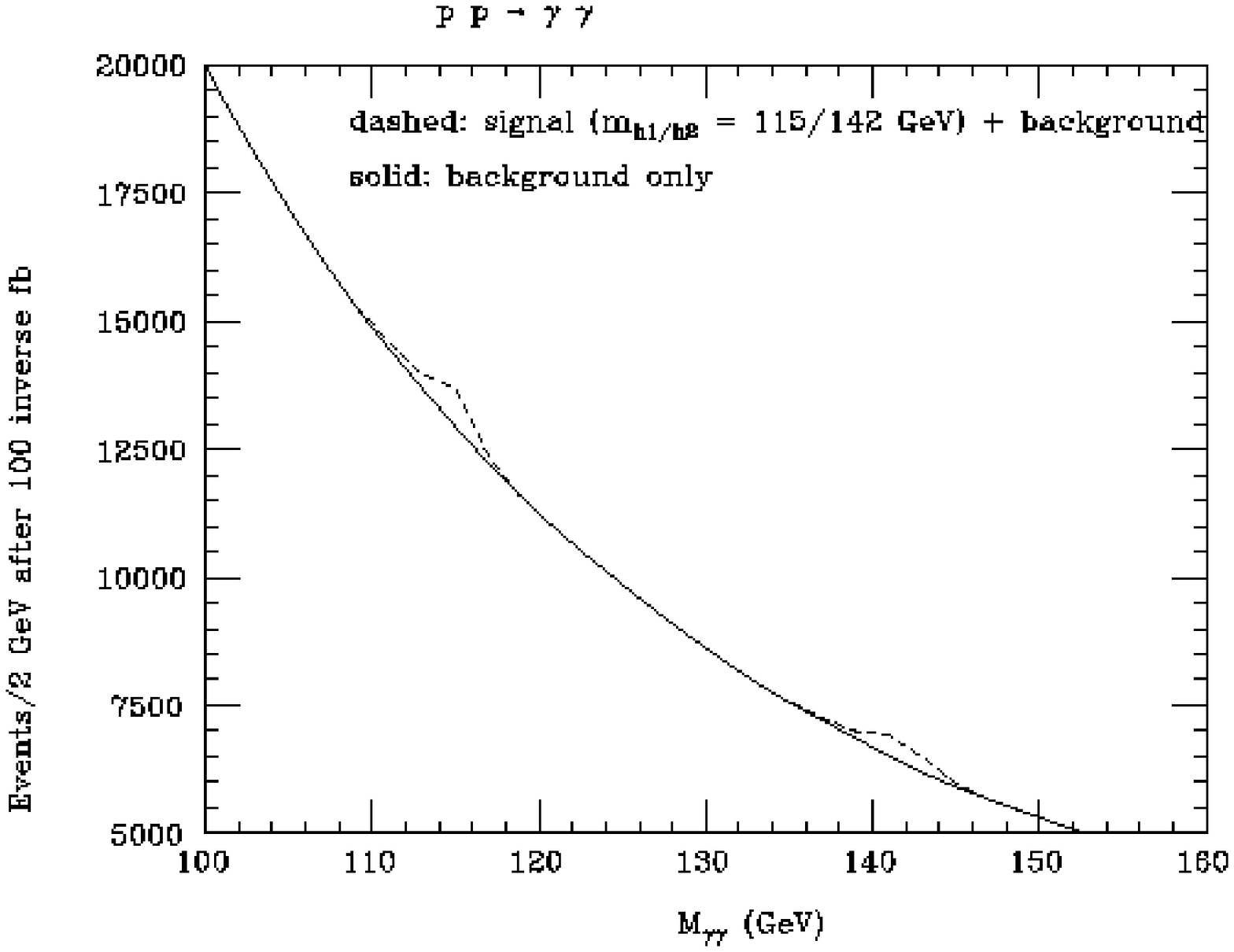}\\[-1.5cm]
\hspace*{3.5truecm}\includegraphics[scale=0.35,angle=90]{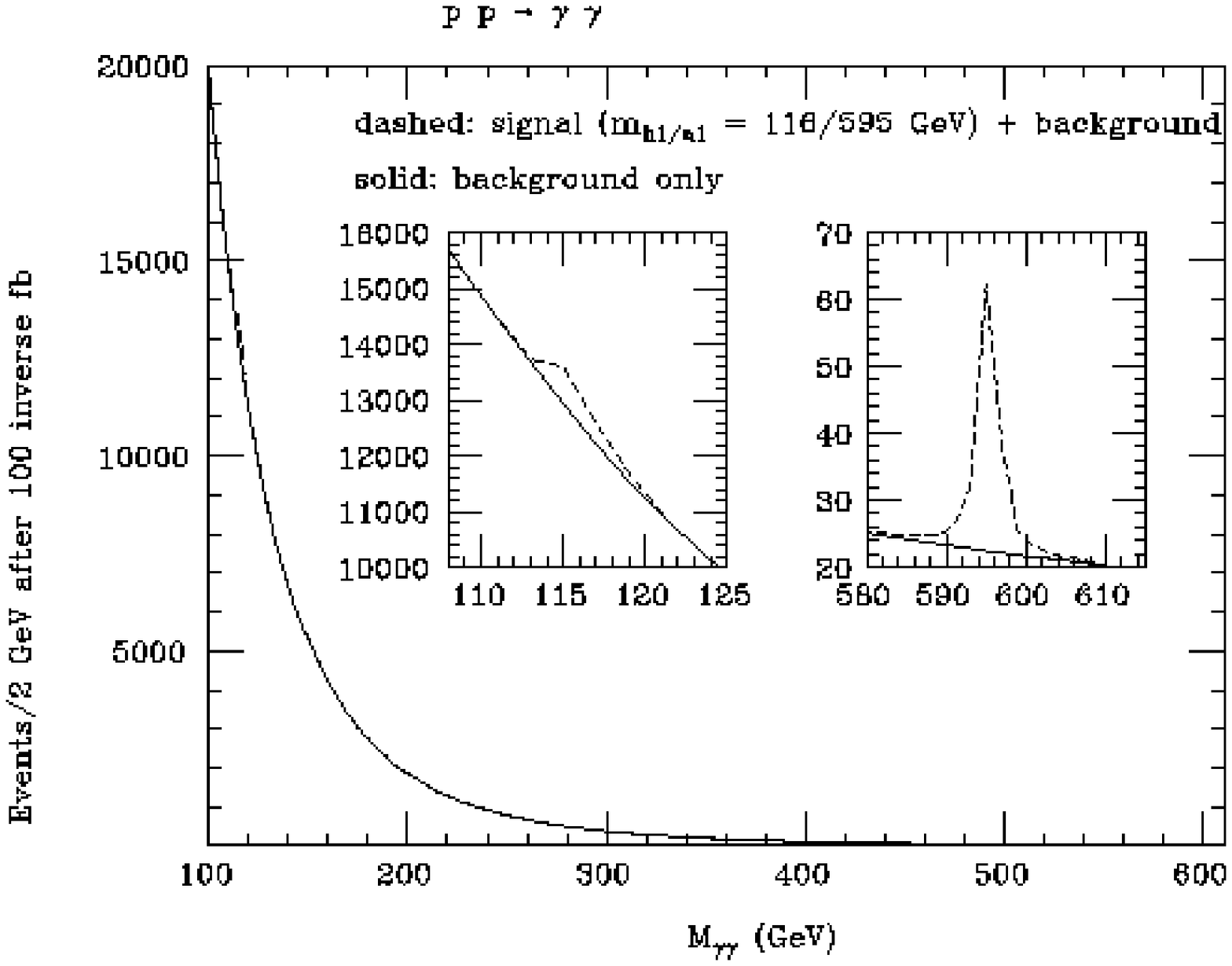}\\
\caption{The differential distribution in invariant mass of the di-photon pair after the cuts in $p_T^\gamma$ and $\eta^\gamma$ mentioned
in the text, for 100 fb$^{-1}$ of luminosity, in the case of the background (solid) and the sum of signal and background (dashed), for 
the example points 1.--3. described in the text (from top to bottom, in correspondence).}
\end{figure}


\begin{thebibliography}{99}

\bibitem{MSSMrev} For reviews see, e.g.:
H.E. Haber and G.L. Kane, Phys. Rep. \textbf{117} (1985) 75;
P. Fayet and S. Ferrara, Phys. Rep. \textbf{32} (1977) 249; 
H.P. Nilles, Phys. Rep. \textbf{110} (1984) 1; 
R. Barbieri, Riv. Nuov. Cim. \textbf{11} (1988) 1.

\bibitem{VHiggs}J.F. Gunion and H.E. Haber, Nucl. Phys. B \textbf{272} (1986) 1; (E) hep-ph/9301205.

\bibitem{muscale}J.E. Kim and H.P. Nilles, Phys. Lett. B \textbf{138} (1984) 150.

\bibitem{NMSSM} 
P. Fayet, Nucl. Phys. B \textbf{90} (1975) 104; Phys. Lett. B \textbf{64} (1976) 159; Phys. Lett. B \textbf{69} (1977) 489 and Phys. Lett. B \textbf{84} (1979) 416;
H.P. Nilles, M. Srednicki and D. Wyler, Phys. Lett. B \textbf{120} (1983) 346; 
J.M. Frere, D.R. Jones and S. Raby, Nucl. Phys. B \textbf{222} (1983) 11; 
J.P. Derendinger and C.A. Savoy, Nucl. Phys. B \textbf{237} (1984) 307;
K. Inoue, A. Komatsu and S. Takeshita, Prog. Theor. Phys \textbf{68} (1982) 927; (E) \emph{ibid.} \textbf{70} (1983) 330;
M. Dine, W. Fischler and M. Srednicki, Phys. Lett. B \textbf{104} (1981) 199; 
A.I. Veselov, M.I. Vysotsky and K.A. Ter-Martirosian, Sov. Phys. JETP \textbf{63} (1986) 489 [Zh. Eksp. Teor. Fiz. \textbf{90} (1986) 838];
B.R. Greene and P.J. Miron, Phys. Lett. B \textbf{168} (1986) 26;
J.R. Ellis, J.F. Gunion, H.E. Haber, L. Roszkowski and F. Zwirner, Phys. Rev. D \textbf{39}  (1989) 844; 
M. Drees, Int. J. Mod. Phys. A \textbf{4}  (1989) 3635;
U. Ellwanger, M. Rausch de Traubenberg and C.A. Savoy, Phys. Lett. B \textbf{315} (1993) 331;
P.N. Pandita, Z. Phys. C \textbf{59} (1993) 575;
S.F. King and P.L. White, Phys. Rev. D \textbf{52}  (1995) 4183.

\bibitem{PQ}R.D. Peccei and H.R. Quinn, Phys. Rev. Lett. \textbf{38} (1977) 1440 and Phys. Rev. D \textbf{16} (1977) 1792.

\bibitem{Z3}M. Cvetic, D.A. Demir and L. Everett, Phys. Rev. D \textbf{56} (1997) 2861;
D.A. Demir and L. Everett, Phys. Rev. D \textbf{69} (2004) 015008;
T. Han, P. Langacker and B. McElrath, Phys. Rev. D \textbf{70} (2004) 115006;
S.F.~King, S.~Moretti and R.~Nevzorov, Phys. Lett. B {\bf 634} (2006) 278, 
Phys. Rev. D {\bf 73} (2006) 035009 and hep-ph/0601269 
(and references therein).

\bibitem{slightly} 
D.J.~Miller and R.~Nevzorov, hep-ph/0309143 and hep-ph/0411275;
D.J.~Miller, S.~Moretti and R.~Nevzorov, hep-ph/0501139.


 
\bibitem{DW} H.P. Nilles, M. Srednicki, and D. Wyler, Phys. Lett. B \textbf{124} (1983) 337;
A.B. Lahanas, Phys. Lett. B \textbf{124} (1983) 341;
U. Ellwanger, Phys. Lett. B \textbf{133} (1983) 187;
J. Bagger and E. Poppitz,  Phys. Rev. Lett. \textbf{71} (1993) 2380;
J. Bagger, E. Poppitz and L. Randall,  Nucl. Phys. B \textbf{426} (1994) 3;
V. Jain, Phys. Lett. B \textbf{351} (1995) 481;
S.A. Abel, Nucl. Phys. B \textbf{480} (1996) 55;
C.F. Kolda, S. Pokorski and N. Polonsky, Phys. Rev. Lett. \textbf{80} (1998) 5263;
S.A. Abel, S. Sarkar and P.L. White, Nucl. Phys. B \textbf{454} (1995) 663.
  
\bibitem{KT} C. Panagiotakopoulos and K. Tamvakis, Phys. Lett. B \textbf{446} (1999) 224.

\bibitem{other-non-minimal} 
C. Panagiotakopoulos and K. Tamvakis, Phys. Lett. B \textbf{469} (1999) 145;
C. Panagiotakopoulos and A. Pilaftsis,  Phys. Rev. D \textbf{63} (2001) 055003;
A. Dedes, C. Hugonie, S. Moretti and K. Tamvakis, Phys. Rev. D \textbf{63} (2001) 055009

\bibitem{DM} A. Menon, D.E. Morrissey and C.E.M. Wagner, Phys. Rev. D \textbf{70} (2004) 035005;
V. Barger, P. Langacker and H.-S. Lee, Phys. Lett. B \textbf{630} (2005) 85; 
J.F. Gunion, D. Hooper and B. McElrath, hep/ph0509024;
G. Belanger, F. Boudjema, C. Hugonie, A. Pukhov and A. Semenov, JCAP \textbf{0509} (2005) 001.

\bibitem{baryon1}
M. Bastero-Gil, C. Hugonie, S.F. King, D.P. Roy and S. Vempati, Phys. Lett. B \textbf{489} (2000) 359; 
S.W.~Ham, S.K.~Oh, E.J.~Yoo, C.M.~Kim and D.~Son,  Phys. Rev. D \textbf{70} (2004) 075001;
K. Funakubo, S. Tao and F. Toyoda, Prog. Theor. Phys. {\bf 114} (2005) 369.

\bibitem{baryon2} M. Carena, M. Quiros, M. Seco and C.E.M. Wagner, Nucl. Phys. B \textbf{650} (2003) 24; 
T. Konstandin, T. Prokopec, M.G. Schmidt and M. Seco,  Nucl. Phys. B {\bf 738} (2006) 1.

\bibitem{MNZ}
D.J. Miller, R. Nevzorov and P.M. Zerwas, Nucl. Phys. B \textbf{681} (2004) 3.

\bibitem{upper}  M. Masip, R. Mu${\tilde{\rm n}}$oz-Tapia and A. Pomarol,
Phys. Rev. D {\bf 57} (1998) 5340.

\bibitem{Cyril} U.~Ellwanger and C.~Hugonie,
Eur.\ Phys.\ J.\ C {\bf 25} (2002) 297.

\bibitem{NoLoseMSSM}J. Dai, J.F. Gunion, R. Vega, Phys. Lett. B \textbf{315}  (1993) 355 and Phys. Lett. B \textbf{345} (1995)  29;
J.R. Espinosa, J.F. Gunion, Phys. Rev. Lett. \textbf{82}  (1999) 1084.

\bibitem{NoLoseNMSSM1} U.~Ellwanger, J.F.~Gunion and C.~Hugonie, hep-ph/0111179;
U. Ellwanger, J.F. Gunion, C. Hugonie and S. Moretti, hep-ph/0305109 and hep-ph/0401228;
D.J. Miller and S. Moretti, hep-ph/0403137.

\bibitem{dirk} D. Zerwas and S. Baffioni, private communication;
S. Baffioni, talk presented at ``GdR Supersym\'etrie 2004, 
5-7 July 2004, Clermont-Ferrand, France.

\bibitem{cNMSSM}C. Hugonie and S. Moretti, hep-ph/0110241.

\bibitem{NMHDECAY}U. Ellwanger, J.F. Gunion and C. Hugonie, JHEP {\bf 0502} (2005) 066.

\bibitem{NoLoseNMSSM2} U.~Ellwanger, J.F.~Gunion and C.~Hugonie, JHEP {\bf 0507} (2005) 041.

\bibitem{WJSZK} Z.~Kunszt, S.~Moretti and W.J.~Stirling,
Z.\ Phys.\ C {\bf 74} (1997) 479.

\bibitem{abdel}A. Djouadi, hep-ph/0503173.

\bibitem{Jacobs} V. Buscher and K. Jacobs, Int. J. Mod. Phys. A {\bf 20} (2005) 2523.

\bibitem{ATLASTDR} ATLAS Collaboration, Technical Proposal  CERN/LHCC/94-43, LHCC/P2, 15 December 1994.

\bibitem{Erice} S. Munir, talk given at the `International School of Subnuclear Physics, 43rd Course', 
Erice, Italy, August 29 -- Sept. 7, 2005, to be published in the proceedings, preprint SHEP-05-37,
October 2005.

\bibitem{NMHDECAY2} U. Ellwanger and C. Hugonie, hep-ph/0508022. 

\bibitem{Alliance} G. Belanger, F. Boudjema and S. Moretti, in preparation.

\bibitem{HERWIG} S.~Moretti, K.~Odagiri, P.~Richardson, M.~H.~Seymour and B.~R.~Webber,
JHEP {\bf 0204} (2002)  028;
G.~Corcella {\it et al.}, JHEP {\bf 0101} (2001)  010.

\bibitem{preparation}  S. Moretti and S. Munir, in preparation.


\end{thebibliography}
\end{document}